\documentclass[twocolumn]{aastex631}
\patchcmd{\linenumbers}
  {\advance\linenumber\@ne}
  {}{}{}
\usepackage{float}

\usepackage{amsfonts}
\usepackage{booktabs}
\let\tablenum\relax
\usepackage{siunitx}
\DeclareSIUnit\bar{bar}
\usepackage{xcolor}
\usepackage{url, microtype, isomath, mathtools, siunitx, physics}
\usepackage{graphicx}
\usepackage{rotating}
\usepackage{booktabs}
\usepackage[version=4]{mhchem}
\usepackage{multirow}
\usepackage{rotfloat}

\shorttitle{New Surface Spectral Library for Rocky Exoplanets}
\shortauthors{Paragas et al.}
\graphicspath{{./}{}}

\begin{document}

\title{A New Spectral Library for Modeling the Surfaces of Hot, Rocky Exoplanets}

\correspondingauthor{Kimberly Paragas}
\email{kparagas@caltech.edu}

\author[0000-0003-0062-1168]{Kimberly Paragas}
\affiliation{Division of Geological and Planetary Sciences, California Institute of Technology, 1200 E California Blvd, Pasadena, CA 91125, USA}

\author[0000-0002-5375-4725]{Heather A. Knutson}
\affiliation{Division of Geological and Planetary Sciences, California Institute of Technology, 1200 E California Blvd, Pasadena, CA 91125, USA}

\author[0000-0003-2215-8485]{Renyu Hu}
\affiliation{Jet Propulsion Laboratory, 4800 Oak Grove Dr, Pasadena, CA 91101, USA}
\affiliation{Division of Geological and Planetary Sciences, California Institute of Technology, 1200 E California Blvd, Pasadena, CA 91125, USA}

\author[0000-0002-2745-3240]{Bethany L. Ehlmann}
\affiliation{Division of Geological and Planetary Sciences, California Institute of Technology, 1200 E California Blvd, Pasadena, CA 91125, USA}

\author{Giulia Alemanno}
\affiliation{Institute of Planetary Research, German Aerospace Center (DLR), Rutherfordstrasse 2, 12489, Berlin-Adlershof, Germany}

\author{J{\"o}rn Helbert}
\affiliation{Institute of Planetary Research, German Aerospace Center (DLR), Rutherfordstrasse 2, 12489, Berlin-Adlershof, Germany}

\author{Alessandro Maturilli}
\affiliation{Institute of Planetary Research, German Aerospace Center (DLR), Rutherfordstrasse 2, 12489, Berlin-Adlershof, Germany}

\author[0000-0002-0659-1783]{Michael Zhang}
\affiliation{Department of Astronomy \& Astrophysics, University of Chicago, 5640 S Ellis Ave, Chicago, IL 60637, USA}

\author[0000-0003-0971-1709]{Aishwarya Iyer}
\affiliation{NASA Goddard Space Flight Center, 8800 Greenbelt Rd, Greenbelt, MD 20771, USA}

\author{George Rossman}
\affiliation{Division of Geological and Planetary Sciences, California Institute of Technology, 1200 E California Blvd, Pasadena, CA 91125, USA}




\begin{abstract}

\textit{JWST}’s MIRI LRS provides the first opportunity to spectroscopically characterize the surface compositions of close-in terrestrial exoplanets. Models for the bare-rock spectra of these planets often utilize a spectral library from \citet{Hu2012}, which is based on room temperature reflectance measurements of materials that represent archetypes of rocky planet surfaces. Here we present an expanded library that includes hemispherical reflectance measurements for a greater variety of compositions, varying textures (solid slab, coarsely crushed, and fine powder), as well as  high temperature (500-800 K) emissivity measurements for select samples. We incorporate this new library into version 6.3 of the retrieval package \texttt{PLATON} and use it to show that surfaces with similar compositions can have widely varying albedos and surface temperatures. We additionally demonstrate that changing the texture of a material can significantly alter its albedo, making albedo a poor proxy for surface composition. We identify key spectral features -- the \SI{5.6}{\micro\meter} olivine feature, the transparency feature, the Si-O stretching feature, and the Christiansen feature -- that indicate silicate abundance and surface texture. We quantify the number of JWST observations needed to detect these features in the spectrum of the most favorable super-Earth target, LHS 3844 b, and revisit the interpretation of its \textit{Spitzer} photometry. Lastly, we show that temperature-dependent changes in spectral features are likely undetectable at the precision of current exoplanet observations. Our results illustrate the importance of spectroscopically-resolved thermal emission measurements, as distinct from surface albedo constraints, for characterizing the surface compositions of hot, rocky exoplanets. 

\end{abstract}

\section{Introduction} \label{sec:intro}
There has been a revolution in our understanding of the properties of small rocky exoplanets over the last decade, mostly driven by results from large space-based transit surveys including \textit{Kepler} and the \textit{Transiting Exoplanet Survey Satellite (TESS)}. When combined with ground-based radial velocity (RV) mass measurements, these transit surveys indicate that nearly all planets with radii less than $R_p < 1.5R_\oplus$ (hereafter referred to as `super-Earths' or `sub-Earths', depending on their radii) have high bulk densities that are well-matched by terrestrial planet models \citep[e.g.,][]{Weiss2014, Rogers2015, Wolfgang2015, Dai2019,Neil&Rogers2020, Luque&Palle2022, Brinkman2024a, Brinkman2024b}. The ongoing \textit{TESS} survey has been particularly effective in identifying new terrestrial planet candidates orbiting nearby low-mass stars; these systems are the most observationally favorable targets for detailed characterization studies \citep[e.g.,][]{Wordsworth2022, TRAPPIST-1JWSTCommunityInitiative2023}. Thanks to the unprecedented capabilities of \textit{JWST}, we can now spectroscopically characterize the surface and atmospheric properties of these rocky exoplanets for the first time \citep{Zhang2024, Hu2024, WeinerMansfield2024, Xue2024}.

Many of the rocky exoplanets detected to date orbit relatively close to their host stars, and as a result are expected to be tidally locked. It is an open question as to whether these small, close-in planets can retain high mean molecular weight atmospheres, or whether they are instead bare rocks analogous to Mercury. Although transmission spectroscopy should in principle allow us to search for evidence of atmospheres on rocky exoplanets, attempts to do so using \textit{JWST} have all resulted in nondetections to date \citep[e.g.,][]{Lim2023, Lustig-Yaeger2023, May2023, Moran2023}.  However, by measuring their dayside brightness temperatures and corresponding day-night temperature gradients, we can use the presence (or absence) of day-night heat transport to indirectly constrain the thickness of their atmospheres \citep{Seager2009}. 
There are currently nine close-in rocky planets with dayside temperature measurements from \textit{Spitzer} or \textit{JWST}, all of which have a cumulative XUV irradiation \citep{Zahnle2017} similar to or greater than that of Mercury. Ten of these planets have hot dayside temperatures indicating 
they have little to no atmosphere \citep{Kreidberg2019, Crossfield2022, Zieba2022,Greene2023, Zieba2023, Gressier2024, Wachiraphan2024, WeinerMansfield2024, Xue2024, Zhang2024}, while one planet \citep[55 Cnc e;][]{Hu2024} appears to have a thick \ce{CO2} and/or \ce{CO}-rich atmosphere.  
55 Cnc e's dayside is hot enough to melt the rocky surface, and it is therefore likely that its escaping atmosphere is continually replenished by outgassing from this region \citep[e.g.,][]{Gaillard2022}. 


The apparent lack of thick atmospheres on most close-in rocky exoplanets means we can use observations of their thermal emission spectra to constrain the compositions of their rocky surfaces. This was first proposed as a method to characterize airless rocky exoplanets in \cite{Hu2012}, inspired by the legacy of studies characterizing rocky surfaces in the Solar System through reflection and thermal emission spectroscopy. Such data are important because knowledge of surface composition provides constraints on a planet's bulk composition and its history of differentiation, volcanism, and weathering. In this study, Hu et al. proposed that broadband near-infrared albedo measurements can discriminate between a subset of surface types, and that types of silicate surfaces on rocky exoplanets can be spectroscopically distinguished via measurements of the prominent mid-infrared Si-O feature \citep[\SIrange{7}{13}{\micro\meter};][]{Hu2012}. 

\begin{figure*}[ht!]
    \centering
    \includegraphics[width=\textwidth]{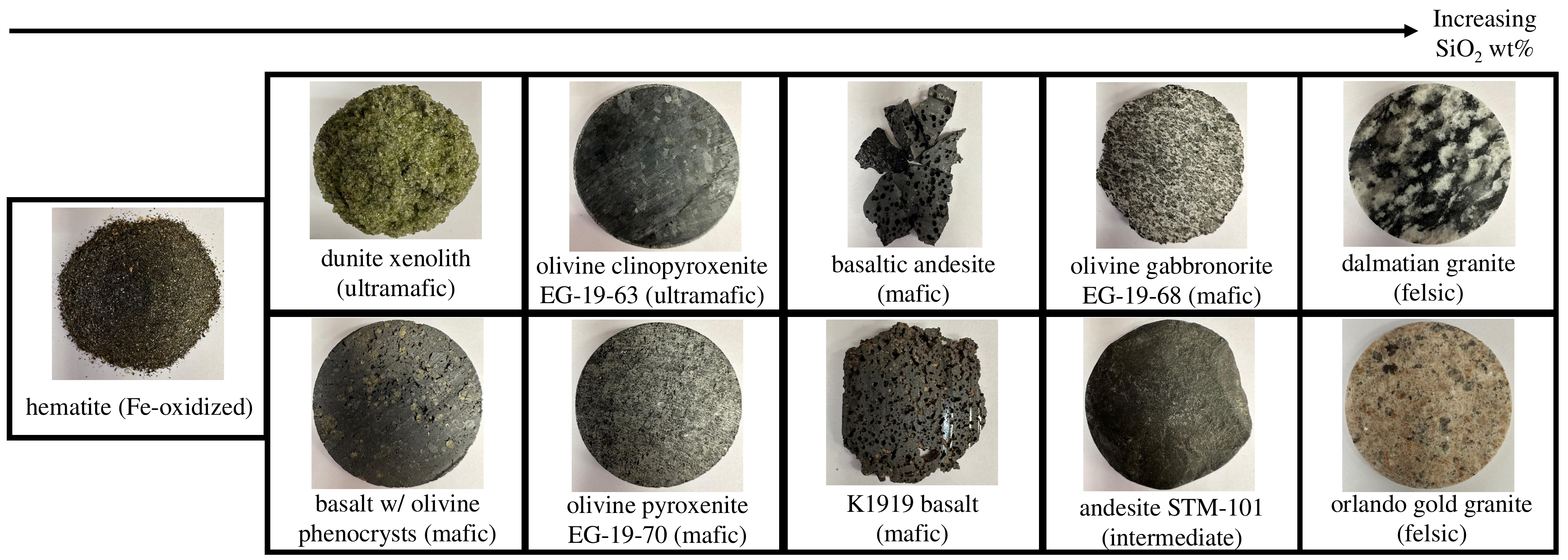}
    \caption{Representative images of of the sample types used in this study.  We show images of the slabs (solid samples) where available. These slabs were cut to a diameter of approximately \SI{45}{\milli\meter} and a thickness of approximately \SI{3}{\milli\meter} as described in \S\ref{sec:sample_selection}. For very porous and/or smaller sample pieces (K1919 basalt, basaltic andesite), we assembled individual fragments to cover an equivalently sized area. For samples without solid slabs (hematitite, dunite xenolith) we show the crushed material (grain sizes between 500~\unit{\micro\meter}$-$1~\unit{\milli\meter}) instead.}
    \label{fig:rock gallery}
\end{figure*} 

We can access these wavelengths observationally with \textit{JWST}'s MIRI Low Resolution Spectroscopy (LRS) mode; this is the first instrument sensitive enough to measure the thermal emission spectra of rocky exoplanets. There are currently four rocky exoplanets with published MIRI LRS mid-infrared emission spectra, including LTT 1445 A b \citep{Wachiraphan2024}, GL 486 b \citep{WeinerMansfield2024}, GJ 1132 b \citep{Xue2024} and GJ 367 b \citep{Zhang2024}. All four of these emission spectra are consistent with a blackbody (i.e., no statistically significant spectral features were detected), and these studies instead used their measurements to place constraints on the planets' Bond albedos of $\leq0.08-0.2$. There are also two additional rocky planets with $T_\mathrm{eq} < 1200$~\unit{\kelvin} (planets likely to still have at least partially solid daysides) with upcoming or archival \textit{JWST} emission spectroscopy using MIRI/LRS: LHS 3844 b (GO 1846, PI Kreidberg) and TOI-4481 b (GO 4931, PI Mansfield). In this study, we focus on the benchmark super-Earth LHS 3844 b \citep{Vanderspek2019}, as it is the most favorable targets for thermal emission spectroscopy and has more extensive \textit{JWST} observations than any of the other planets on this list. In addition to the three MIRI LRS observations mentioned above, its near-infrared (\SIrange{3}{5}{\micro\meter}) phase curve was also measured with NIRSpec in Cycle 2 (GO 3615, PI Zieba). Although results from both programs are still pending, \citet{Kreidberg2019} previously found that this planet's high \textit{Spitzer} \SI{4.5}{\micro\meter} dayside brightness temperature was best matched by an airless, low-albedo surface similar to those inferred for the three other planets with published MIRI LRS observations. 

In this study, we aim to develop improved surface spectral models for interpreting the emerging ensemble of rocky exoplanet emission spectra from \textit{JWST}. Unlike models of atmospheric thermal emission, where the same underlying set of opacities can be used to model arbitrary mixtures of gases over a wide range of temperatures and pressures, the thermal emission spectra of rocky surfaces are not well-approximated by models utilizing libraries of mineral spectra and most remote sensing studies of solar system objects therefore rely on laboratory measurements of analogous materials to interpret their observations. In the exoplanet community, most thermal emission models \citep[e.g.,][]{Kreidberg2019, Mansfield2019, Whittaker2022, Wachiraphan2024, WeinerMansfield2024, Xue2024, Zhang2024} for rocky exoplanets are currently derived from the small spectral library presented in \citet{Hu2012}. Recently there have been new efforts to use the RELAB database for modeling rocky exoplanet observations \citep[e.g.,][]{Hammond2024}, although this study simulated broadband photometric (MIRI F1500W and F1280W) observations. Additionally, \cite{First2024} recently published a new library of 15 basaltic samples with varying degrees of aqueous alteration and used this library to show that aqueous alteration features may be detectable for the most ideal target, LHS 3844 b, if present. However, this seems unlikely given LHS 3844 b's measured dayside temperature of $1040\pm40$~K \citep{Kreidberg2019}. 

Silicate rocks are variable compositionally, classified by both bulk chemistry and mineralogy. Nearly identical bulk chemistries can have different spectral properties because they have different mineral assemblages, which depend on factors like temperature, pressure, oxidation state, and volatile content. Infrared spectra are also very sensitive to the surface texture (i.e., bulk rock vs. regolith), but the spectral library presented in \citet{Hu2012} only included powdered samples. Although rocky exoplanets initially form with solid surfaces, impact bombardment will comminute the surface over time if the planet lacks a thick atmosphere, as is observed in the solar system \citep[e.g.,][]{Hapke2001, Domingue2014}.
We know very little about the frequency of volcanic resurfacing events on close-in rocky exoplanets \citep[e.g.,][]{Lichtenberg2024}, and these planets therefore might exhibit a wide range of surface textures. 

There are extensive libraries of reflectance and emission spectral libraries of rocks at wavelengths relevant for interpreting \textit{JWST} data \citep[e.g., the Salisbury library; USGS library; ASU thermal emission spectral library; the RELAB database;][]{Christensen2001, Milliken2021}. Howevever, the vast majority of these data were taken at room temperature, while the airless rocky exoplanets observed in emission by \textit{JWST} have surfaces that are substantially hotter than room temperature ($>800-1000$~K). The spectra of minerals and rocks are known to change with increasing temperature \citep[e.g.,][]{Helbert2013, Ferrari2020} and with thermal gradient effects under lunar-like conditions \citep[in vacuum with heating from above and below;][]{DonaldsonHanna2017}. It is therefore important to quantify the importance of these temperature-dependent effects for interpreting exoplanet observations. Recently, \citet{Fortin2024} presented the first high temperature spectral library aimed at understanding \textit{JWST} observations of lava worlds, which consists of infrared emissivity spectra of eight synthetic glass samples measured between $\sim$\SIrange{1050}{1650}{\kelvin}. Here we examine a complementary set of natural materials spanning a wider range of textures and grain sizes than those considered in this study, and consider a slightly lower temperature range more appropriate for solid (versus partially melted) surfaces.   


In this study, we characterize the reflectance and emission spectra of a representative set of primarily igneous samples with minimal aqueous alteration, as these rock types are the most relevant proxies for the surfaces of hot rocky exoplanets. We begin by measuring hemispherical reflectance and emission spectra for all samples, including 
three different textures: a solid slab, coarsely crushed grains, and a fine powder. We then measure the high temperature emissivities for a subset of our samples between $500-800$~K to explore temperature-dependent changes in their spectral features and quantify the importance of these effects for exoplanet surface models.  We incorporate our expanded spectral library into the open-source PLATON modeling code \citep{Zhang2019, Zhang2020}, and use these models to explore the degeneracies between dayside albedo and surface properties.  We then discuss specific mid-infrared spectral features that can provide unique diagnostics of surface composition and texture, and quantify their detectability with \textit{JWST}.
Lastly, we use this new modeling framework to interpret published \textit{Spitzer} \SI{4.5}{\micro\meter} photometry of the super-Earth LHS 3844 b, and to make predictions for upcoming \textit{JWST} near- and mid-infrared spectroscopy of its dayside emission spectrum.

\section{Experimental Methods}

\subsection{Sample Selection and Preparation}\label{sec:sample_selection}
We began by collecting samples of igneous rocks spanning the ultramafic, mafic, intermediate, and felsic categories, as well as one Fe-oxidized sample. We assigned our silicate rocks into each of these categories based on their \ce{SiO2} content: $\leq 45\%$ (ultramafic), $45-52\%$ (mafic), $52-63\%$ (intermediate), and $\geq 63\%$ (felsic). On Earth, ultramafic rocks 
form from the solidification of melts similar in composition to the mantle, while on Mars they are thought to have formed as primary crust \citep{Elkins-Tanton2005}. They also likely make up some of the lavas on Mercury \citep{Charlier2013}. Mafic rocks comprise the majority of Earth's crust, the lunar mare, and the surface of Mars. Intermediate rocks are more rare than mafic or felsic rocks on Earth, and are typically associated with subduction zones. Felsic rocks are 
commonly found on Earth as a result of plate tectonics. Lastly, Fe-oxidized materials have been found in abundance on Mars \citep{Ehlmann2014}. 

We measured the reflectance of all samples with an ASD FieldSpec (FS3 Max) Spectrometer (see \S\ref{subsec:asd_meas}) between \SIrange{0.35}{2.5}{\micro\meter} to search for signatures of weathering, such as aqueous alteration. We eliminated any samples with strong absorption features at $1.9$~\unit{\micro\meter} and $2.2-2.3$~\unit{\micro\meter}, which are typical indicators of aqueous alteration, and signify formation of secondary mineral phases due to the fact that we do not expect water to be present on the hot, rocky exoplanets that are the best targets for potential surface characterization. On Earth, nearly all natural rocks have some incipient weathering from interaction with water, so we accepted samples with small 1.9 and 3.0 \unit{\micro\meter} features due to water that are masked in later plots. In a recent work focused on the detectability of surface composition for rocky exoplanets, \citet{First2024} presented emissivity spectra of 15 basaltic samples that were not screened for aqueous alteration. However, the hot, rocky exoplanets ($T>700~\unit{\kelvin}$), including LHS 3844 b, that offer the best opportunity for surface characterization are not expected to retain the volatile inventory necessary for producing such alteration \citep{Hu2012}. We also prioritized the inclusion of samples spanning a range in \ce{SiO2} abundances. Our final spectral library includes 11 different rock types ranging from ultramafic to felsic rocks, more than doubling the number of samples in these compositional classes from \cite{Hu2012}. We have one Fe-oxidized sample: hematite; two ultramafic samples: dunite xenolith and olivine clinopyroxenite (EG-19-63); five mafic samples: olivine pyroxenite (EG-19-70), basaltic andesite, a fine-grained basalt from the Kilauea 1919 flow (K1919), a basalt with olivine phenocrysts, and olivine gabbronorite (EG-19-68); one more feldspar-rich sample of intermediate composition: andesite (STM-101); two felsic samples: dalmatian granite and orlando gold granite.  Representative images of all of the samples can be found in Figure~\ref{fig:rock gallery}. 

Our samples encompass four of the broad categories of rock types described in \citet{Hu2012}--ultramafic, basaltic, granitoid, and Fe-oxidized.  Unlike their study, we did not include any clay or ice-rich silicate samples, as the rocky exoplanets accessible to \textit{JWST} emission spectroscopy are uniformly too hot for liquid surface water or water ice. We also did not include any non-oxidized metal-rich materials; while these may be relevant for other planetary systems, they would have required special rock synthesis and measurement under reducing atmosphere conditions and were therefore beyond the scope of this study. We were also unable to obtain a suitable feldspathic lunar highlands-like anorthosite sample as in \citet{Hu2012}. 

We processed our selected samples into three textures: a solid slab, crushed, and powdered. For the slab, we drilled \SI{45}{\milli\meter} diameter cores and then sliced them into \SI{3}{\milli\meter} disks. The dimensions of these disks were dictated by the dimensions of the crucible used for our high-temperature emissivity measurements as described in \S\ref{sec:high_temp_meas}. For the crushed and powdered samples, we ground rock pieces down using a shatterbox and then sieved the resulting materials to include grains of \SI{500}{\micro\meter} to \SI{1}{\milli\meter} (crushed) and \SIrange{25}{63}{\micro\meter} (powdered) in size. The dunite xenolith was too crumbly to properly core and slice, and the hematite had already been crushed.  We therefore do not have slab samples for either of these materials.

\begin{figure*}[ht!]
    \centering
    \includegraphics[width=\textwidth]{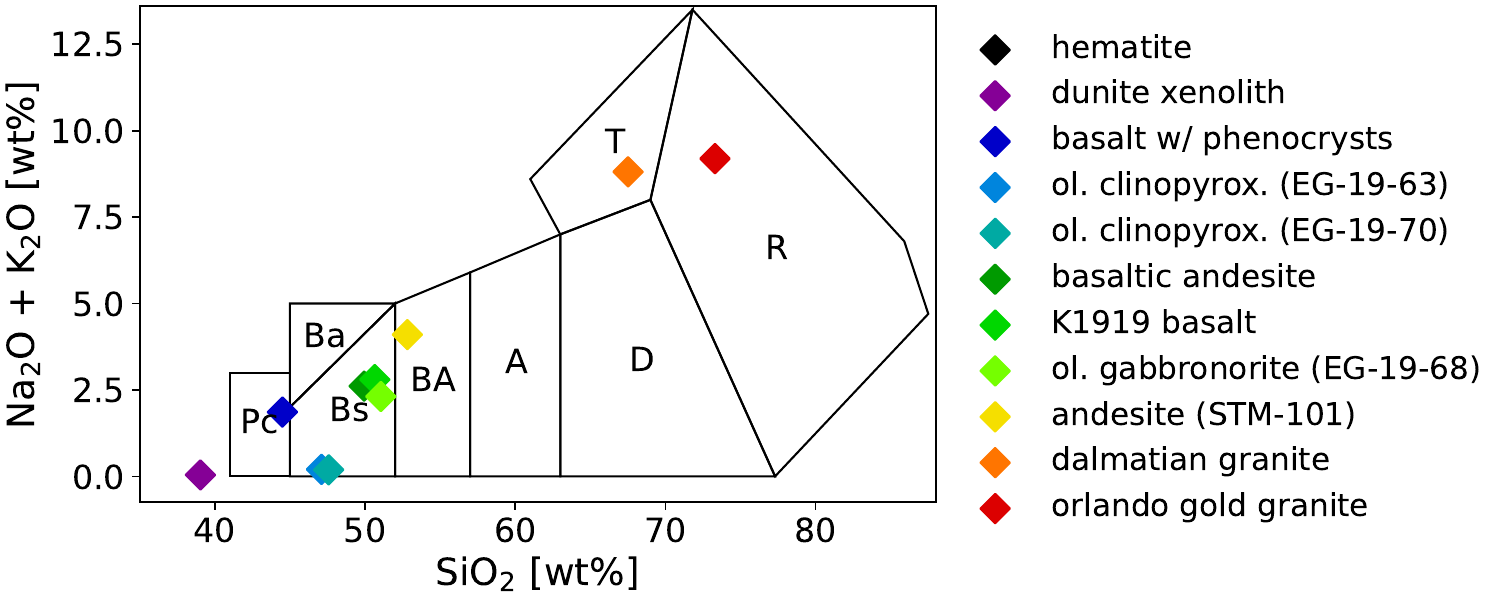}
    \caption{Total Alkali Silica (TAS) diagram of all igneous samples in this study.  The labels Pc, Bs, Ba, BA, A, D, T, and R correspond to picrite basalt, basalt, trachybasalt, basaltic andesite, andesite, dacite, trachyte, and rhyolite, respectively. Though this compositional space is intended for igneous rocks, we include the dunite xenolith sample in order to provide visual comparison of its chemical abundances relative to the entire sample set. The hematite sample has an \ce{SiO2} wt\% of 2.15 (contaminant of silicate in the sample), and we do not expand the bounds of this plot to include it.}
    \label{fig:TAS}
\end{figure*}

As a final step, we characterized the detailed properties of each rock sample by submitting them to Actlabs for whole rock chemical analysis (4B - Lithium Metaborate/Tetraborate Fusion - ICP) and X-ray diffraction mineral identification (quantitative). For the igneous rocks (all except the hematite), we show their \ce{Na2O} + \ce{K2O} and \ce{SiO2} wt\% in a Total Alkali Silica (TAS) diagram in Figure~\ref{fig:TAS}. The detailed chemistry and mineral characterization results are in Table~\ref{table:chemistry} and Table~\ref{table:mineralogy}, respectively, in the Appendix. 

\subsection{Laboratory Measurements}\label{sec:lab_meas}
All hemispherical reflectance and high temperature emissivity measurements presented in this work were performed in evacuated conditions at the German Aerospace Center (Deutsches Zentrum fuer Luft- und Raumfahrt, or DLR) Institute for Planetary Research in Berlin, Germany. Additionally, the near-infrared spectra of our samples were measured using an ASD FieldSpec in the Ehlmann lab at Caltech.

\subsubsection{Hemispherical Reflectance}
Hemispherical reflectance \citep[the total fraction of light scattered into all directions in the upward going hemisphere by a surface illuminated from above by a collimated light source;][]{Hapke2012} measurements for each sample and its respective textures were measured with an incidence angle of $13$\unit{\degree} in evacuated conditions (\SI{0.7}{\milli\bar}) at room temperature between \SIrange{0.35}{1.11}{\micro\meter} (visible; VIS) and \SIrange{1.5}{25}{\micro\meter} (mid-infrared; MIR) using a Bruker Vertex80V FTIR spectrometer. The VIS reflectance measurements were acquired using a spectralon-coated integrating sphere, an Si-Diode detector, a UV-VIS-NIR-\ce{CaF2} beamsplitter, and a VIS lamp internal to the spectrometer, and referenced against measurements of a spectralon-coated plate. The MIR reflectance measurements were acquired using a gold-coated integrating sphere, an MCT cooled down detector, a KBr beamsplitter, and a MIR lamp internal to the spectrometer, and referenced against measurements of a gold-coated plate. Our spectra have a spectral resolution of \SI{4}{cm^{-1}} and are created by averaging 500 multiple interferograms collected by the instrument to assure high values of signal-to noise ratio in our data \citep{Maturilli2006}. More detailed information on the laboratory set-up at the DLR can be found in \citet{Maturilli2019}.

\subsubsection{ASD Reflectance Measurements}\label{subsec:asd_meas}
We used an Analytical Spectral Devices (ASD) FieldSpec (FS3 Max) Spectrometer to measure relative reflectance of all of our samples between \SIrange{0.35}{2.5}{\micro\meter}; this effectively fills in a gap in near-infrared coverage of the Bruker FTIR spectrometry, and also allowed us to eliminate samples from consideration that showed signs of significant aqueous alteration. We used the ASD spectrometer in the ``contact probe" configuration for which the light source is internal from an overhead halogen bulb and light is collected by a fiber at an 18-degree angle. For every measurement, we placed the contact probe flush with the material. We calibrated the reflectances with a spectralon reference which we remeasured every ten minutes. We corrected for the properties of the spectralon in a process performed internally in the ASD RS3 spectral processing exporter. We collected and averaged measurements for both pre-heated and post-heated spectra. For pre-heated spectra, we averaged 100 measurements each of the standard, dark current, and sample. For post-heated spectra, we averaged 50 measurements of each type.

\subsection{Constructing a Combined Optical to Mid-IR Hemispherical Reflectance Spectrum}
The Bruker software automatically processes the data to produce calibrated hemispherical reflectance measurements in each spectral range, which we used as the starting point for our analysis. We then multiplied the MIR hemispherical reflectances by a factor of 0.95 to take into account the real reflecting power of the gold standard used as a reference in order to express them as absolute reflectances. We also removed a feature from each individual VIS spectrum caused by a laser signal between \SIrange{0.6326}{0.6336}{\micro\meter}. To remove this laser signal, we calculated the median reflectance values for the left (between \SIrange{0.632}{0.6326}{\micro\meter}) and right (between \SIrange{0.6336}{0.6342}{\micro\meter}) continuum regions directly surrounding the feature, and replaced the reflectance values corresponding to the laser feature with the average of these two values. 

\begin{figure*}[ht!]
    \centering
    \includegraphics[width=0.87\textwidth]{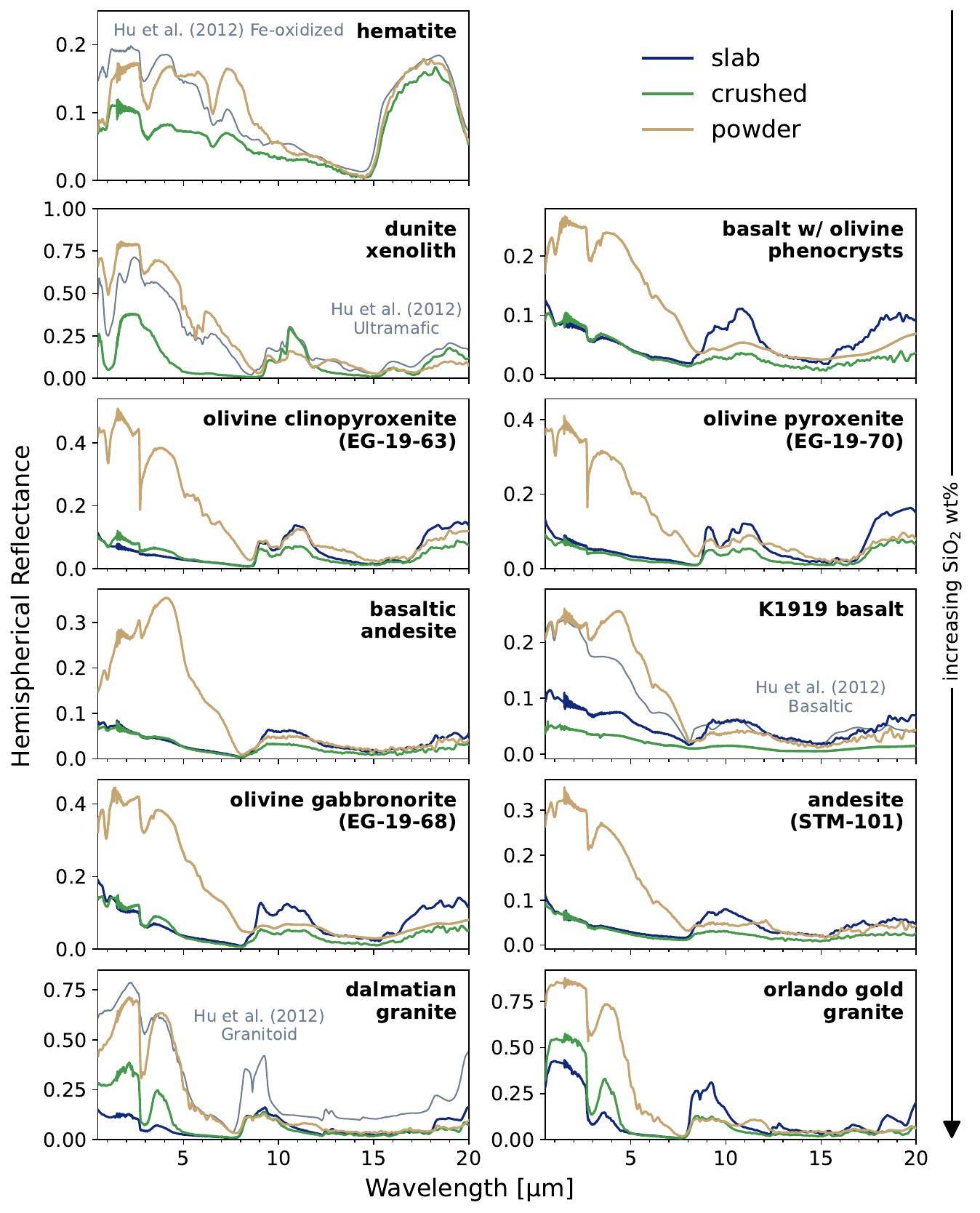}
    \caption{Final calibrated hemispherical reflectance measurements. The name of each sample is given in the top right corner of each panel, with the color of the curves in each panel corresponding to the texture of the sample. Whenever available, we obtained measurements for three textures: slab (dark blue), crushed (green; $500$~\unit{\micro\meter} to $1$~\unit{\milli\meter} grains), and powder (gold; \SIrange{25}{63}{\micro\meter} grains). We also show hemispherical reflectance spectra for the each of the materials described in \citet{Hu2012} overplotted as grey curves on the panels corresponding to the closest compositional match in our new sample library for comparison. The feldspathic and basaltic samples in this older study were derived from measurements of rock powders with diameters less than $200$~\unit{\micro\meter}, while the remaining three samples (Fe-oxidized, ultramafic, and granitoid) were calculated using Hapke radiative transfer modeling of the endmember minerals.}
    \label{fig:all hemi refl plots}
\end{figure*}

Our models require a single continuous reflectance spectrum spanning optical to mid-infrared wavelengths. However, there is a gap in wavelength coverage of our optical and infrared hemispherical reflectance spectra between \SIrange{1.11}{1.5}{\micro\meter}. We also found that there were vertical offsets between the measured optical and infrared reflectances for some samples. This is not surprising, as the VIS and MIR measurements were taken with different detectors, and each sample had to be removed and then replaced between measurements. While the compositions of the crushed and powdered samples were relatively uniform within the sample cup, some slabs exhibited larger inhomogeneities and the resulting spectrum was therefore sensitive to their orientation within the instrument. The optical and infrared data for the crushed and powdered samples generally matched up well, while the largest offsets between the measured optical and infrared reflectances were associated with the slab samples. We used data from the ASD FieldSpec Spectrometer, which measures spectra between \SI{0.35}{\micro\meter} and \SI{2.5}{\micro\meter}, to fill in the gap in the DLR Bruker measurements. We took the VIS measurements as our starting point and normalized the ASD spectra to the these measurements by matching their reflectance values at the longest VIS wavelength of \SI{1.11}{\micro\meter}. We then normalized the MIR hemispherical reflectance measurements to match the renormalized ASD spectrum between \SIrange{1.5001}{1.5219}{\micro\meter}, and filled the gap with the ASD spectrum values in the relevant wavelength range. The normalization values and a figure showcasing this procedure are in Appendix~\ref{appendix: f}. The resulting smoothed and calibrated hemispherical reflectances are shown in Figure~\ref{fig:all hemi refl plots}.

\subsection{High Temperature Emissivity Measurements}\label{sec:high_temp_meas}
We measured high temperature emissivities for a subset of ten samples from our spectral library, where the total number of samples was dictated by the available time in the laboratory. 
For these measurements, we prioritized the five most different rocks in terms of their \ce{SiO2} content: basalt with olivine phenocrysts (slab), dunite xenolith (crushed, no slab available), orlando gold granite (slab), K1919 basalt (slab), and the andesite STM-101 (slab). 
We also included the olivine pyroxenite EG-19-70 (slab) to have an additional mafic sample. Lastly, we measured the hematite (crushed, no slab available) as it was our only Fe-oxidized sample. We also measured powdered textures for three samples to investigate how texture affected temperature-dependent changes in \ce{SiO2} features. These powders were the K1919 basalt, dunite xenolith, and the orlando gold granite.

\begin{figure}[ht!]
    \centering
    \includegraphics[width=0.47\textwidth]{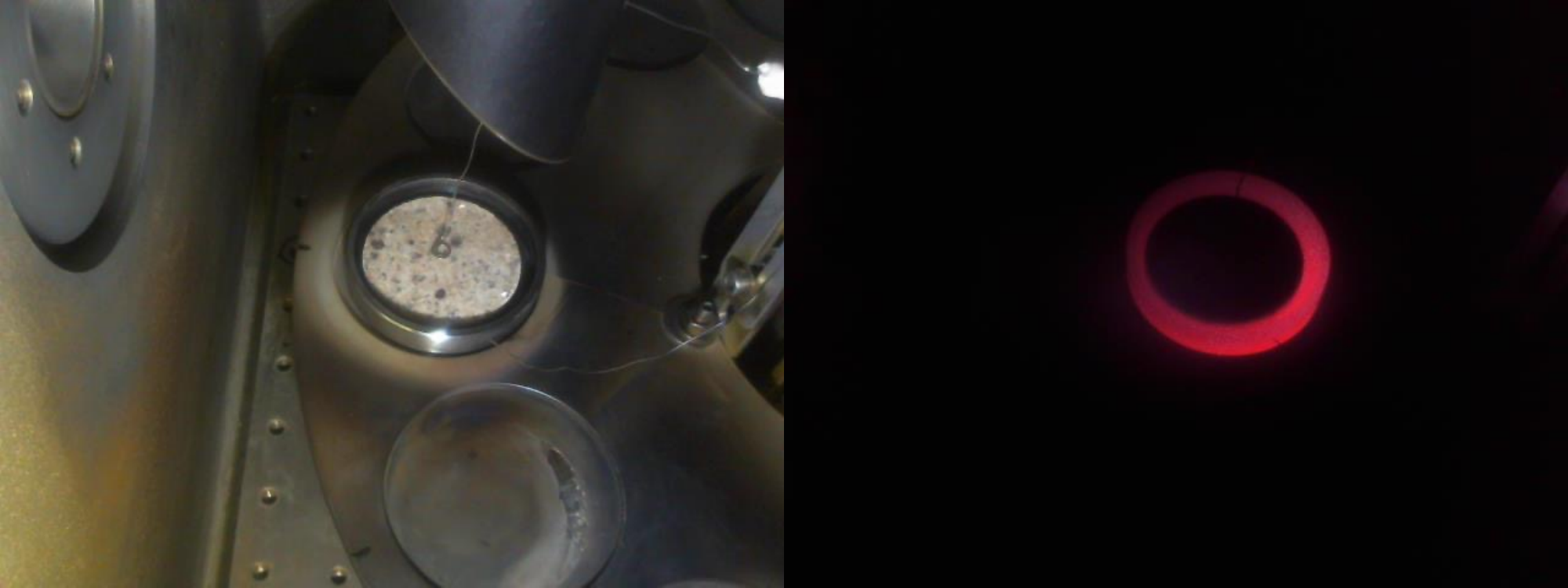}
    \caption{Photos of the orlando gold granite slab sample taken with the webcam mounted in the emissivity chamber. The picture on the left was taken at room temperature with the light inside the chamber turned on. The picture on the right was taken at $\sim$\SI{684}{\kelvin} with the light turned off to accentuate the glowing sample cup.}
    \label{fig:emissivity chamber}
\end{figure}

We obtained our emissivity measurements using an in-house designed and built emissivity chamber, externally coupled to a Bruker Vertex80V FTIR spectrometer with an MCT cooled detector and a KBr beamsplitter as described in \cite{Maturilli2006}. We placed each sample in a steel sample cup and evacuated the chamber to pressures below $\sim$\SI{0.1}{mbar}. We used induction to heat the samples and monitored the temperatures of both the sample surface and the rim of the sample holder using high sensitive thermal sensors. We also monitored the samples visually throughout the heating process using a webcam mounted in the chamber. Representative photos of the orlando gold granite slab are shown in Figure~\ref{fig:emissivity chamber}. We measured the emission spectrum of each sample at the following approximate temperature steps: \SI{500}{K}, \SI{600}{K}, \SI{700}{K}, \SI{800}{K} and \SI{830}{K}. The last measurement was taken at the maximum temperature reachable for the used measurement setup and samples. Our emissivity spectra have the same spectral resolution as the reflectance spectra of \SI{4}{cm^{-1}}. We also measured the room-temperature hemispherical reflectances of each sample before and after heating in order to check for alterations caused by the heating process. There are several differences between our experimental set-up and the one described in \citet{Fortin2024}. First, \citet{Fortin2024} used a furnace to heat the sample, making the surrounding areas the same temperature as the sample, whereas we heated our samples via induction from the sample cup. Second, they filled the chamber with dry air and streamed argon into the furnace and onto the sample whereas our measurements were obtained in a vacuum. Lastly, the spectral range of their measurements is modestly larger (\SIrange{2.5}{20}{\micro\meter}) than our spectral range (\SIrange{4}{18}{\micro\meter}).

We converted our laboratory emission spectra to wavelength-dependent emissivity measurements using a graphite reference target as described in \citet{Maturilli2006} and \citet{Maturilli2014} with some updates. We first constructed a library of reference measurements by heating the graphite target and recording its radiance using the same experimental setup and spanning a similar range of temperatures to our science data. These reference measurements have been proven to be extremely stable and repeatable in emissivity \citep{Maturilli2014}. 
Next, we fit the measured intensity spectrum $I_\mathrm{sample,3}(\lambda)$ for the middle (third of five) temperature step by optimizing for the temperature $T$ of the sample with the following model $M$ for the intensity,
\begin{equation}
    M(\lambda) = I_\mathrm{ref}(T,\lambda) [1 - (s*r_{h}(\lambda)],
\end{equation}
where $I_\mathrm{ref}(T,\lambda)$ is the reference wavelength-dependent intensity at a specific temperature, $r_{h}(\lambda)$ is the room temperature hemispherical reflectance, and $s$ is a scale factor ranging from 0 to 1 to account for the shallower measured spectral feature depths in the high temperature emissivity data (see \S\ref{sec:highT_discussion} for a more detailed discussion). After determining the best-fit temperature $T_\mathrm{best}$, we calculated the calibrated emissivity $\varepsilon_\mathrm{sample,3}(\lambda)$ as
\begin{equation}
    \varepsilon_\mathrm{sample,3}(\lambda) = \varepsilon_\mathrm{ref}\frac{I_\mathrm{sample,3}(\lambda)}{I_\mathrm{ref}(T_\mathrm{best},\lambda)},
\end{equation}
where $\varepsilon_\mathrm{ref}$ is the measured emissivity of the reference blackbody material (0.975 at all wavelengths). Finally, we normalized this calibrated high-temperature emissivity measurement to the room temperature derived emissivity $1 - r_h(\lambda)$ by subtracting a vertical offset between the maxima of the two. For the silicate samples, we used the emissivity maximum between \SIrange{7}{10}{\micro\meter} (the Christiansen Feature) and for the hematite, we instead used the emissivity maximum between \SIrange{14}{16}{\micro\meter} to determine the vertical offsets. This final step closely follows the default procedure for generating calibrated emissivities in previous studies using the same experimental set-up \citep{Maturilli2006,Maturilli2014}. 

We calculated the calibrated emissivities $\varepsilon_\mathrm{sample, n}(\lambda)$ for each of the remaining four temperature steps  using the same equations as above. However, instead of optimizing for the temperature that best-matched our measured intensity spectrum $I_\mathrm{sample,n}(\lambda)$ we instead optimized for the temperature that provided the closest match between $\varepsilon_\mathrm{sample,n}(\lambda)$ and $\varepsilon_\mathrm{sample,3}(\lambda)$. 
We adopted this approach because we found that when we optimized the blackbody temperatures to match the measured intensity spectra $I_\mathrm{sample,n}(\lambda)$ we obtained inconsistent slopes in the calibrated emissivities $\varepsilon_\mathrm{sample,n}(\lambda)$ across the five temperature steps. We speculate that this may be due to the presence of modest amounts of background flux from the cup and/or chamber, which are not explicitly included in our model and can bias our fitted temperatures. We found that optimizing for the best match between the emissivity spectra resulted in significantly better consistency in the measured slopes, and confirmed that the best-fit temperatures obtained using this method are within $\sim10$~\unit{\kelvin} of the values obtained by directly fitting the measured intensity spectra $I_\mathrm{sample,n}(\lambda)$.

\begin{figure*}[ht!]
    \centering
    \includegraphics[width=0.9\textwidth]{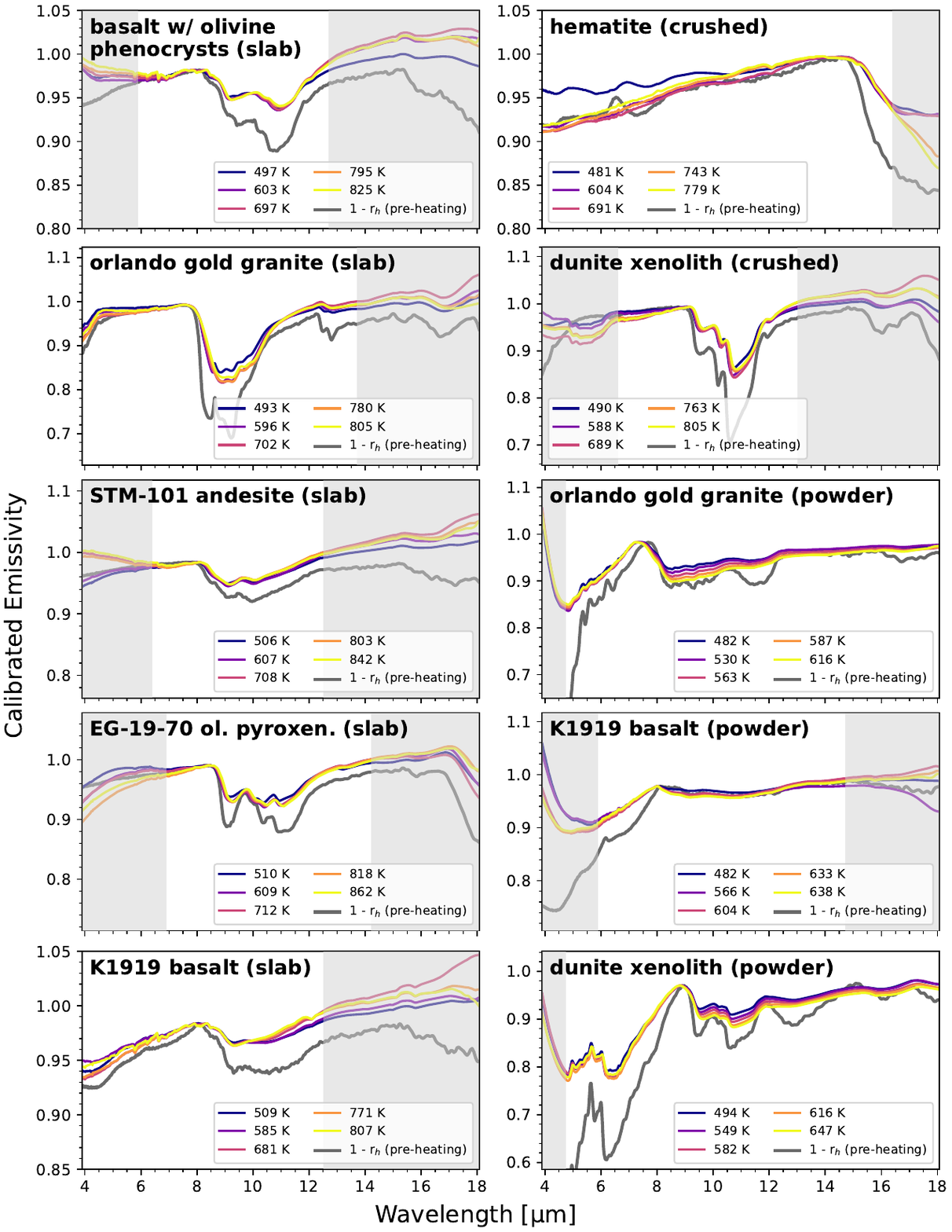}
    \caption{High temperature emissivity measurements of heated samples with their pre-heating directional emissivities (solid gray line) calculated from our room temperature hemispherical reflectance measurements from Fig. \ref{fig:all hemi refl plots} plotted for comparison. The temperatures for each individual emissivity spectrum are derived from our best-fit blackbody calibrator model and are indicated by the color in the legend. Regions of the spectrum that are affected by contaminating flux from the metal cup holding the sample, or where the calibrated spectral shape appears to be inconsistent across temperature bins, are marked with shaded gray rectangles. It is important to note that the absolute emissivity for these measurements is not very meaningful, only its variation with wavelength.}
    \label{fig:all emi comp plots}
\end{figure*}

The final calibrated high temperature emissivity measurements for each of our samples are shown in Figure~\ref{fig:all emi comp plots} with their best-fit sample temperature in the legend of each axes. We find that some samples exhibit a sharp rise in emissivity towards the shortest wavelengths, including all three powdered samples; this is likely due to contaminating thermal emission from the metal cup used to heat the sample (see Fig. \ref{fig:emissivity chamber}). We also see some discrepancies at the longest wavelengths, where the flux from our sample was rapidly decreasing; this may be due to either a contaminating flux source or to a bias in the fitted temperature created by short-wavelength flux contamination. We mask both regions in our plots using gray rectangles. Although high temperature emissivity measurements of the empty sample cup might have allowed for better calibration of any contaminating flux, we unfortunately were unable to obtain such measurements during our time in the laboratory.

\section{Modeling the Bare-Rock Surfaces of Exoplanets} \label{sec:modeling}
We incorporated the resulting library of reflectance measurements into the open-source \texttt{PLATON} code \citep{Zhang2019, Zhang2020,Zhang2024}, which was originally developed to model the transmission and emission spectra of gas giant exoplanets with hydrogen-dominated atmospheres. We adapted this code to model the surface emission spectra of hot, rocky exoplanets observed during secondary eclipse for which the measurable is the secondary eclipse depth, or the planet-to-star flux ratio. In this updated version, users can choose surfaces from the new library presented in this work, the spectral library of \citet{Hu2012}, or input their own hemispherical reflectance measurements. These changes are available as part of PLATON release 6.3 \footnote{\url{https://github.com/ideasrule/platon}}.

We constructed our dayside thermal emission models by following a similar method to the one described in \citet{Hu2012}. Unlike the models in that paper, which divided the planet's day side into a grid and solved for the local equilibrium temperature at each grid location, our models assume a single global dayside temperature. In order to solve for the predicted dayside temperature, we require the directional-hemispherical reflectance and the hemispheric emissivity. We calculated both from hemispherical reflectance measurements as described below. We do not do this with our emissivity measurements because they only cover the mid-infrared, and the visible and near-infrared wavelengths are the most influential in determining the dayside temperature. First, we used the hemispherical reflectance measurements to calculate the single scattering albedo $\omega$ and the directional emissivity $\varepsilon_d$ for each texture of each sample that was measured. Next, we followed the simplest version of the procedure described in \citet{Hapke2012}, which makes the following assumptions: (1) particles scatter isotropically, (2) the opposition effect is negligible for integral reflectances, (3) roughness is negligible (observations not taken close to the limb or terminator), and (4) the porosity parameter is set to unity (as it is extremely difficult to determine). We used the following expressions from \citet{Hapke2012} to estimate the single scattering albedo $\omega$ as a function of wavelength:
\begin{equation}
\gamma(r_{h}) = \frac{1 - r_{h}}{1 + 2\mu_{0}r_{h}}
\end{equation}
\begin{equation}
\omega = 1 - \gamma^{2}
\end{equation}
In this equation, $\gamma$ is the albedo factor, $r_{h}$ is the hemispherical reflectance, and $\mu_{0}$ is the cosine of the incidence angle $i$. For our hemispherical reflectance measurements, we have $i=13^{\circ}$, and thus, $\mu_{0} = 0.97$. 

Using this single scattering albedo, we calculated the hemispheric emissivity $\varepsilon_{h}$, defined as the hemispherical average of the directional emissivity, with the following expressions from \citet{Hapke2012}
\begin{equation}
\varepsilon_{h} \simeq \left(1 - r_{0}\right)\left(1 + \frac{r_{0}}{6}\right)
\end{equation}
\begin{equation} 
r_{0} = \frac{1 - \gamma}{1 + \gamma}
\end{equation}
in which $r_{0}$ is the diffusive reflectance.  This quantity is distinct from the directional emissivity $\varepsilon_{d}$, which is given by $\varepsilon_{d} = 1 - r_h$.

We follow a similar method to that of \citet{Hu2012} to solve for the dayside surface temperature that satisfies energy balance, with one exception. Instead of using a geometric albedo defined by the radiance coefficient, we instead use the directional and hemispherical emissivities calculated from our hemispherical reflectance measurements as inputs in our energy balance equation:
\begin{equation} 
f \int\varepsilon_{d}F_\mathrm{inc}d\lambda = \int\varepsilon_{h} B_{\lambda}(T_\mathrm{surf})d\lambda
\end{equation}
where $f$ is a 1D correction factor to account for how the absorbed stellar energy is redistributed across the planet (see Appendix~\ref{appendix: f}), $F_\mathrm{inc}$ is the incident energy on the planet and $B_\lambda$ is the Planck function. The integration is over the wavelengths corresponding to our hemispherical reflectance measurements between \SIrange{0.5}{23.5}{\micro\meter}. We repeated our calculations using the alternative energy balance equation given in \citet{Hu2012} and found that the differences between emission spectra calculated using these two approaches were negligible. We opt for this version of the expression because we are using hemispherical reflectances rather than bidirectional reflectances \citep[as done in][]{Hu2012}. This distinction means that we cannot compute geometric albedos for our samples without making assumptions about their particle phase functions, and therefore we do not model the temperature in patches across the dayside as a function of longitude and latitude. Nonetheless, our 1D approach provides almost equivalent spectra to the 2D models. We therefore utilize the expression above moving forward. 

Our 1D models assume that the dayside surface of the planet has a uniform temperature $T_\mathrm{surf}$, and that no heat is redistributed to the night side. In this limit, $f = 1/2$ is needed to ensure that the energy radiated by the planet is equal to the energy it absorbs from the star. However, in reality, the day sides of tidally locked airless bodies should have a temperature gradient that reflects the spatially varying incident flux at the surface. This means that the regions of the planet near the substellar point will be hotter than the regions near the terminator. When we view the dayside of the planet face-on, the region near the substellar point contributes more to the hemisphere-integrated flux than the cooler terminator regions, and setting $f = 1/2$ causes us to underestimate the total measured flux for this viewing geometry. \citet{Hansen2008} showed that in this case, $f = 2/3$ produces a better approximation for the measured dayside flux. 

We determine the optimal value of $f$ for each surface type in  \citet{Hu2012} by solving for the value of $f$ that provides the best match between our single temperature model and the predicted emission spectrum calculated using \citet{Hu2012}'s 2D models, which account for the temperature gradient across the dayside as a function of longitude and latitude. In this calculation, we assumed that the host was a main sequence star with an effective temperature of \SI{3000}{\kelvin}; this should be reasonably representative of most rocky planet systems that are accessible to \textit{JWST}. We found that in all cases, our best-fit $f$ values varied between 1/2 and 2/3. Although the value for $f$ also has a weak dependence on equilibrium temperature (colder planets prefer slightly smaller $f$ values), we opted to use a single $f$ value for each sample calculated for a planet with $T_\mathrm{eq} = 1000$~\unit{\kelvin}. We found that our fitted $f$ values were an approximately linear function of the median hemispherical reflectance value for each sample type over a wavelength range corresponding to 85\% of the integrated stellar flux for a \SI{3000}{\kelvin} M star. We then used a linear fit to these points to estimate $f$ values for each of the new samples in this study. Although several of our new samples have median reflectances greater than the most reflective surface types in \citet{Hu2012}, we find that the maximum $f$ value is still greater than 0.5. A more detailed discussion of this calculation can be found in Appendix~\ref{appendix: f}.

Once we know the optimal value of $f$ for a given surface type and texture, we can then solve for the dayside temperature of the planet and calculate the corresponding emitted $F_\mathrm{e}$ and reflected flux $F_\mathrm{r}$ as a function of wavelength. For the emitted flux, we multiply the directional emissivity by $\pi$ times the Planck function, and for the reflected flux, we multiply the hemispherical reflectance by the stellar flux $F_{*}$ and the ratio of the stellar radius and the semi-major axis. Thus, the total hemisphere-integrated dayside planet flux $F_\mathrm{planet}$ is given by summing the two, i.e.,  
\begin{equation}
F_\mathrm{planet} = \pi\varepsilon_d B_{\lambda} + r_{h} F_{*} \left(R_{*}/a\right)^{2},
\end{equation}
where $R_{*}$ is the stellar radius, and $a$ is the semi-major axis of the planet. To calculate the predicted secondary eclipse depth as a function of wavelength, we then divide the planet flux by the stellar flux and multiply by the planet-star radius ratio squared. 

\begin{figure}[ht!]
    \centering
    \includegraphics[width=0.47\textwidth]{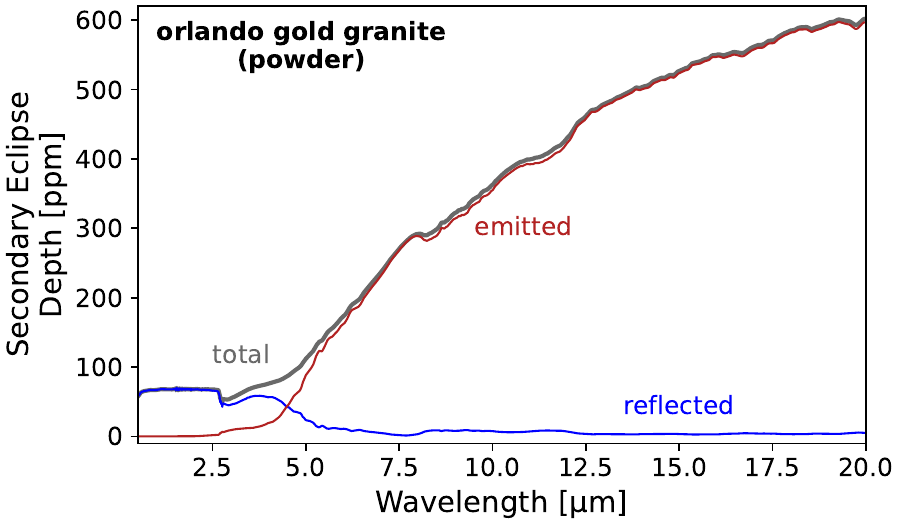}
    \caption{The secondary eclipse depths, or planet-to-star flux ratio, for an LHS 3844 b analog with a bare-rock surface composed of orlando gold granite powder, the most reflective surface and texture combination in our library. These models were generated assuming a blackbody stellar spectrum.}
    \label{fig:refl light}
\end{figure}

Although our final models are dominated by thermal emission at longer wavelengths, they also include reflected light and can therefore be used to interpret optical and near-infared observations. 
Figure~\ref{fig:refl light} shows the predicted wavelength-dependent secondary eclipse depth for a LHS 3844 b analog with a bare-rock surface of orlando gold granite powder. By LHS 3844 b analog we mean that we are using the parameters of the LHS 3844 b system as inputs to our model rather than implying that the surface of the planet is actually granite. The orlando gold granite is the most reflective surface and texture combination in our library, but even for this sample we find that reflected light only dominates for wavelengths shortward of \SI{5}{\micro\meter}. 

\section{Results and Discussion}
The spectra of rocky materials are highly variable and depend on the composition, grain size, and temperature of the material, among other factors \citep[e.g.,][]{Hapke2012, Helbert2013}. In this section, we use our new spectral library to explore how variations in these properties can affect our ability to interpret secondary eclipse observations of hot, bare-rock exoplanets and constrain their surface compositions. It is important to keep in mind that prior to the launch of \textit{JWST}, published measurements of thermal emission from rocky exoplanets were typically obtained in a single broad photometric bandpass. In the absence of spectroscopic information, these studies instead used the measured dayside flux to constrain the planet's dayside surface temperature and corresponding wavelength-integrated surface albedo.  This in turn allowed them to rule out a subset of the possible surface compositions from the spectroscopic library of \citet{Hu2012} \citep[e.g.,][]{Kreidberg2019,Crossfield2022}. This same approach was also used in the first three studies presenting JWST MIRI thermal emission spectroscopy of close-in rocky exoplanets, as their measured spectral shapes were consistent with blackbodies \citep{Zhang2024, WeinerMansfield2024, Xue2024}
In the following sections, we use our newly expanded spectral library to demonstrate two key takeaways for exoplanet observers. First, albedo alone provides relatively weak and degenerate constraints on surface properties. The albedos of samples with similar compositions can vary by up to a factor of two, while the albedos of individual samples can vary by up to a factor of seven depending on their texture. 
Second, we identify some of the strongest spectral features at mid-infared wavelengths that may be detectable with \textit{JWST} emission spectroscopy, and discuss how these features can be used as unique diagnostics of exoplanet surface compositions and textures. We conclude that temperature-dependent changes in the shapes of these spectral features are likely below the noise floor of current exoplanet observations. Finally, we demonstrate the degeneracies between albedo and surface properties with the use of our new spectral library by interpreting published \emph{Spitzer} 4.5 micron photometry of LHS~3844~b \citep{Kreidberg2019}.

\subsection{Albedo Does Not Uniquely Map to Composition}\label{sec:albedo}
\subsubsection{Varying Albedos within Compositional Classes}\label{sec:comp_diversity}
\begin{figure}[ht!]
    \centering
    \includegraphics[width=0.47\textwidth]{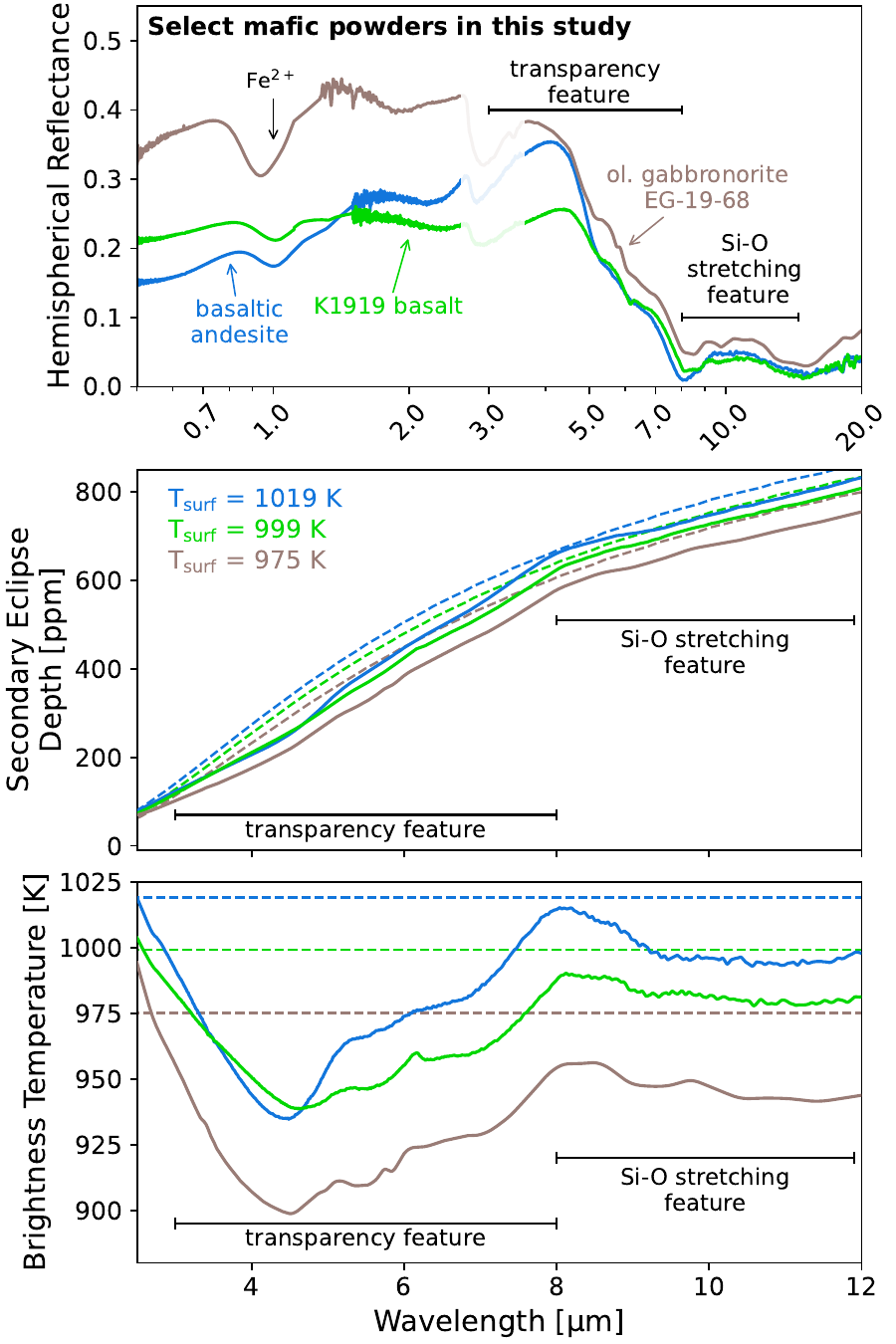}
    \caption{The hemispherical reflectance measurements for a subset of mafic powders (basaltic andesite, K1919 basalt, and olivine gabbronorite EG-19-68) in our study. We label the strong $\sim$\SI{1}{\micro\meter} absorption feature due to octahedrally-coordinated \ce{Fe^{2+}} in olivine and/or pyroxene, the transparency feature characteristic of fine-grained material shortward of \SI{8}{\micro\meter}, and the Si-O stretching feature characteristic of silicate rocks between $\sim$\SI{8}{\micro\meter} and \SI{12}{\micro\meter} with black. The lighter portions of the lines indicate regions containing aqueous alteration-related features in secondary minerals that will likely not be present in surface spectra of hot, rocky exoplanets. In the middle panel, we plot the corresponding predicted wavelength-dependent secondary eclipse depth (this is equivalent to the planet-star flux ratio) with the same color scheme for a LHS 3844~b analog with each surface type. In the lower panel, we plot the brightness temperature of the secondary eclipse depths with the same color scheme. The dashed lines are blackbody curves with the predicted temperatures from the bare-surface models of the corresponding surface.}
    \label{fig:compositional variety}
\end{figure}

The optical hemispherical reflectances and corresponding predicted dayside temperatures can vary significantly within the same compositional class due to differences in mineralogies, grain size, and other sample properties. To illustrate this point, we consider three mafic rocks in our study that occupy the same region in the TAS diagram (Figure~\ref{fig:TAS}) with $\sim$50 \ce{SiO2} wt\% and $\sim$2.5 \ce{Na2O} + \ce{K2O} wt\%: basaltic andesite, K1919 basalt, and olivine gabbronorite (EG-19-68). Controlling for texture by considering only the powdered measurements, we see that the EG-19-68 sample is approximately twice as reflective at wavelengths shortward of \SI{3}{\micro\meter} (see Figure~\ref{fig:compositional variety}). This difference in reflectivity is due to the higher proportion of plagioclase in the EG-19-68 sample (70 wt\% versus 21 and 48 wt\% in the other two samples, see Table \ref{table:mineralogy}). In contrast, the other two samples are noticeably darker due to the greater proportion (35 wt\% for both samples versus 20 wt\% in the olivine gabbronorite) of pyroxene minerals (augite and diopside).

Crucially, we find that the varying hemispherical reflectances of these three mafic powders (see top panel of Figure~\ref{fig:compositional variety}) result in a correspondingly wide range in predicted dayside temperatures over the wavelength range included in the figures (\SIrange{0.5}{20}{\micro\meter}) . We demonstrate this by calculating predicted emission spectra for a planet with the same properties as LHS 3844 b \citep{Vanderspek2019}. We find that a planet with a surface similar to the EG-19-68 powder will have a dayside temperature $20-45$~\unit{\kelvin} cooler than a planet with a surface analogous to the K1919 basalt or basaltic andesite powders. The lower temperature of the powdered EG-19-68 model surface results in a predicted planet-star flux ratio that is $45-85$~ppm lower than the other two surface types at \SI{8}{\micro\meter} (Figure~\ref{fig:compositional variety}). 
We conclude that albedo is an unreliable and degenerate proxy for surface composition, and provide a more detailed demonstration of this in \S\ref{sec:LHS3844b}.

\subsubsection{Surface Albedos Depend on Surface Texture} 
Next, we use our newly expanded library to explore the effect of varying surface texture on our model emission spectra. As shown in Figure~\ref{fig:all hemi refl plots}, changes in texture changes reflectance at near-infrared and optical wavelengths. This is due to increased scattering from interfaces relative to absorption by the material in the near-infrared and increased volume scattering (multiple-scattering) relative to surface scattering (single-scattering) in the mid-infrared, as grain size decreases \citep[e.g.,][]{Mustard1997}. The largest spectral contrast changes are in the visible/near infrared between \SIrange{0.4}{2.5}{\micro\meter} and between \SIrange{4}{7}{\micro\meter} in the so-called transparency region, as discussed in \S\ref{sec:detecting spectral features}. For silicate rocks there is also a decrease in the contrast of the Si-O stretching feature with decreasing grain size \citep{Salisbury1989}, although this is a relatively small effect compared to the increased slope of the transparency feature. 

\begin{figure}[ht!]
    \centering
    \includegraphics[width=0.47\textwidth]{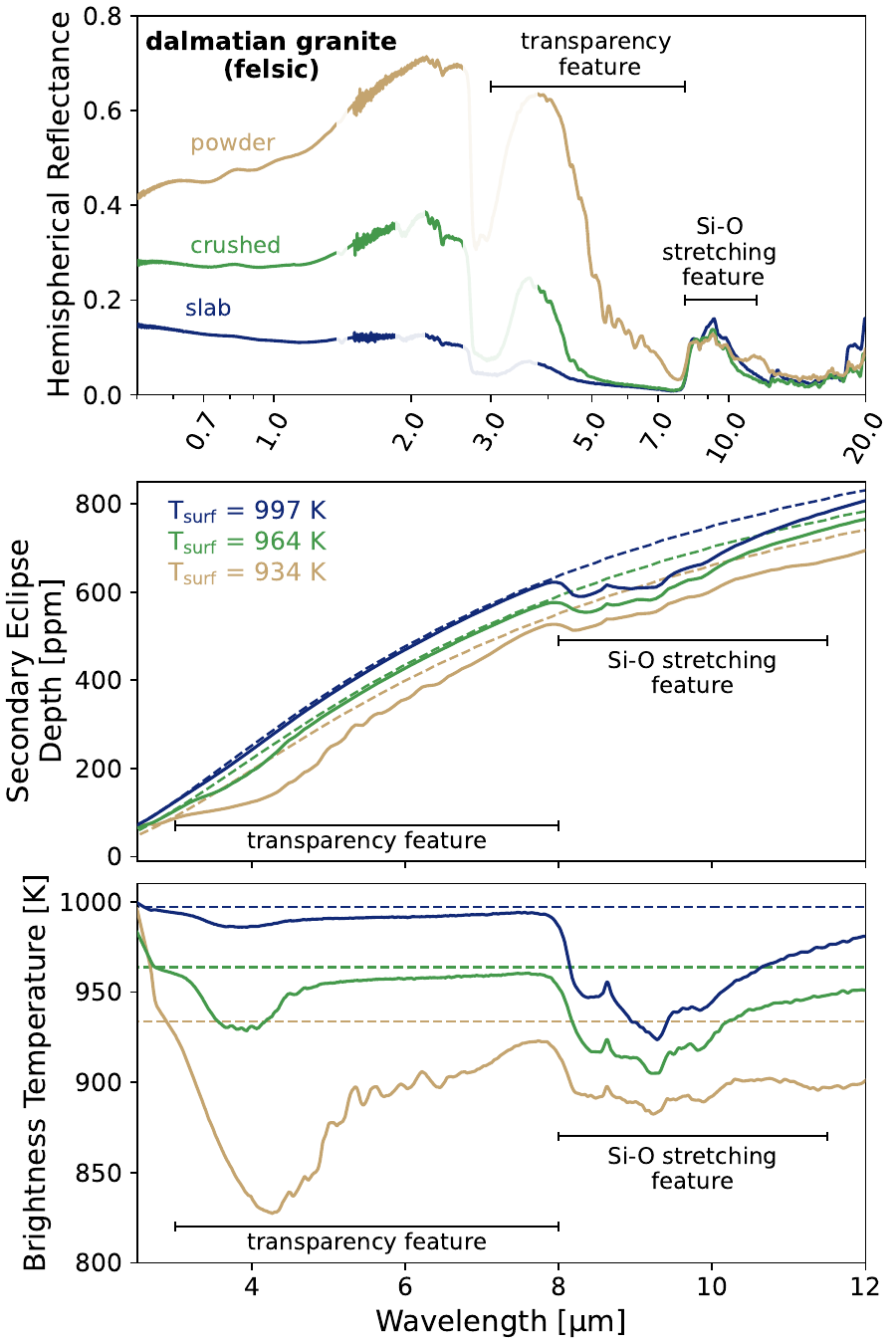}
    \caption{The hemispherical reflectance measurements for the slab, crushed, and powder textures for the dalmatian granite sample. We labeled the transparency feature and the Si-O stretching feature in black. The lighter portions of the lines indicate regions which correspond to alteration features that will likely not be present in the spectra of hot, rocky exoplanets. The middle and lower panels show the secondary eclipse depths and brightness temperature of each texture and corresponding blackbody spectra for an LHS 3844~b analog.}
    \label{fig:textures}
\end{figure}

We use the dalmatian granite sample to illustrate the effect of texture differences on the predicted surface temperature and wavelength-dependent planet-star flux ratio of a LHS 3844~b analogue planet in Figure~\ref{fig:textures}.  We find that the differences in reflectance between the slab and powdered textures for the dalmatian granite sample translate to a \SI{70}{\kelvin} difference in predicted dayside temperature for a LHS~3844~b analog. This temperature difference is similar in magnitude to the temperature difference caused by compositional variations in \S\ref{sec:comp_diversity}. 

\subsection{Prospects for Detecting Spectral Features in the Emission Spectra of Rocky Exoplanets} \label{sec:detecting spectral features}

In Figures~\ref{fig:compositional variety}, \ref{fig:textures}, and \ref{fig:CF depths}, we identify four representative features in the hemispherical reflectance measurements of the ultramafic powders and use them as case studies to demonstrate the feasibility of detecting spectroscopic features in the thermal emission spectra of rocky exoplanets. (we do not include the high contrast \SI{3}{\micro\meter} feature because it is due to \ce{OH}/\ce{H2O} in minerals and is not expected for hot, rocky exoplanets) 

\subsubsection{Ferrous Iron Feature}
The ferrous iron (\ce{Fe^{2+}}) feature near \SI{1}{\micro\meter} is associated with electronic absorptions in Fe-bearing minerals \citep[e.g.,][]{Rossman2019}, is particularly prominent in olivine-rich materials, and is one of the deepest absorption features visible in Fig. \ref{fig:compositional variety}. The depth of this feature is dependent on texture and the band center occurs at slightly different wavelengths in the three ultramafic samples. The shape and band center are known to vary depending on the coordination of the transition metal atom in a mineral structure \citep{Clark1999}, so it is expected that samples with different mineralogical compositions would have different feature shapes. This feature is visible in our predicted dayside emission spectra but has a very small amplitude due to the reduced planet-star flux ratio and correspondingly shallow secondary eclipse depth at near-infrared wavelengths. Unfortunately, its location at \SI{1}{\micro\meter} makes it effectively undetectable from a signal-to-noise perspective for even the most observationally favorable rocky exoplanets. However, this feature would be favorable for detection in the reflected light, if the future Habitable World Observatory finds rocky exoplanets with minimal atmospheres in more widely separated orbits.

\subsubsection{Olivine Feature at \SI{5.6}{\micro\meter}} \label{sec:olivine feature}
The emission feature at \SI{5.6}{\micro\meter} for the dunite xenolith sample (see Figure~\ref{fig:CF depths}) is characteristic of olivine and is the result of the vibrational overtone-combination bands of the Restrahlen bands \citep{Salisbury1991, Bowey2005}. Although the other two ultramafic samples contain some olivine and may show a weak feature at this wavelength in their hemispherical reflectance spectra, this feature is too small to be detected in either of their secondary eclipse depth models. 
We therefore focus on the 
dunite xenolith model as a `best case' scenario for the detectability of this feature.

For LHS 3844 b, we investigated how many secondary eclipses would be necessary in order to detect this \SI{5.6}{\micro\meter} olivine feature using \textit{JWST} MIRI LRS. We simulated the \textit{JWST} observations with realistic uncertainties using \texttt{Pandexo} \citep{Batalha2017} with our modeled thermal emission spectra as inputs, and specified a wavelength resolution of $R=15$. For these tests, we only use the simulated data corresponding to wavelengths between \SI{5}{\micro\meter} and \SI{12.5}{\micro\meter}. As our null hypothesis case, we fit the simulated data with the dunite xenolith powder surface (with the olivine feature removed) scaled by an overall amplitude (the only free parameter). For the detection test, we again fit the spectra with the same surface with the olivine feature removed and fit for the overall amplitude, but include a second amplitude (a second free parameter) that scales the feature at its respective wavelengths. We completed these fits using static nested sampling with the \texttt{dynesty} package \citep{Speagle2020}. We used 300 live points, bound~=~``multi", and sample~=~``rwalk". We repeated these fits on 1000 different random realizations of the \texttt{Pandexo} generated data, and found that at least seven eclipse observations are necessary to detect the olivine feature to greater than $3\sigma$ confidence level as determined by the Bayes factor. 

\subsubsection{Transparency Feature} \label{sec:transparency feature}
The transparency feature is characterized by a peak reflectance at $\sim 4$~\unit{\micro\meter} accompanied by a steep reduction in reflectance longward to $\sim 7$~\unit{\micro\meter}, and is most prominent for smaller grain sizes. As the grain size (divided by the wavelength of interest) and corresponding absorption coefficient decreases, the silicate grains become optically thin, resulting in an increase in multiple scattering \citep{Salisbury1989}. By contrast, solid slabs only exhibit surface scattering in this spectral range, i.e., no multiple scattering takes place \citep{Zhuang2023}. The transparency feature is readily visible for the powdered samples in Fig. \ref{fig:compositional variety} and \ref{fig:textures}. We find that the slope of the feature is quite different across the three ultramafic samples in Fig. \ref{fig:compositional variety}, especially when comparing the dunite xenolith to the other two samples. We speculate that there may be more multiple scattering occurring in the dunite xenolith powder than in the other samples due to its overall higher reflectivity (lower absorptivity), resulting in a steeper transparency feature. 

We expect that this feature should be detectable with broad spectroscopic coverage in the \SIrange{2}{8}{\micro\meter} wavelength range.  We used a similar process as in \S\ref{sec:olivine feature} to determine the number of eclipses necessary to detect the transparency feature. For this test, we selected the dalmatian granite powder (shown in Fig~\ref{fig:textures}) because it has relatively few spectral features in this wavelength range, making it easier to isolate the effect of the transparency feature. We specified a wavelength resolution of $R=10$ for our simulated data to maximize signal-to-noise over the relatively broad wavelength range spanned by this feature. In order to avoid overlap with the Si-O stretching feature, we also limited our fits to simulated data at wavelengths shortward of \SI{8}{\micro\meter}. We fit the simulated data using a grid of models that were interpolated between the solid slab and powder textures of the dalmatian granite sample for these detection tests. This method allowed us to empirically scale the size of the transparency feature, but can also be viewed as equivalent to varying the average grain size. The null hypothesis for this test was a blackbody surface with temperature as a free parameter. To test whether we detect the transparency feature we calculated the Bayes factor for the interpolated grid of models containing the transparency feature compared to the null hypothesis. Since the interpolated models are very similar to each other, we placed a dlogz constraint of 0.5 on the nested sampling routine to keep the sampler from becoming stuck. We changed sample from ``rwalk" to ``unif", but kept bound~=~``multi". We repeated these fits on 1000 different iterations of the \texttt{Pandexo} generated data, and found that at least twenty eclipse observations are necessary to detect the transparency feature to $\geq3\sigma$ confidence level if only using MIRI LRS. Given this large number, we performed this test again but instead used a combination of NIRSpec G395H and MIRI LRS simulate data. We repeated these fits on 1000 different iterations of the \texttt{Pandexo} combined data, and found that with two NIRSpec G395H and three MIRI LRS eclipse observations, the transparency feature can be detected to $4.4\sigma$. This is particularly promising as there are two \textit{JWST} GO programs that have or will measure this exact combination. We also performed this same test on the STM-101 andesite, K1919 basalt, and EG-19-63 olivine clinopyroxenite powders to understand the range in detectability of the transparency feature for each broad category of igneous rock. The different surface types required three/four, three/four, and three/six NIRSpec/MIRI eclipses, respectively. Notably, this may be a means of constraining the surface texture of rocky exoplanets -- fine particulate vs. solid rock -- similar to the way the transparency feature is used in solar system remote sensing, e.g., as a dust vs bedrock discriminator in global Mars mapping \citep[e.g.,][]{Ruff1997}.
 
\subsubsection{Si-O Stretching Feature} \label{sec:SiO stretching feature}
The Si-O stretching feature shown in Figures~\ref{fig:compositional variety} and \ref{fig:textures} also lies in the range covered by MIRI LRS. This feature arises from the vibrational Si-O asymmetric stretch fundamental, and is the strongest diagnostic spectral feature for the presence of silicates \citep{Salisbury1993, Hu2012}. This feature is typically located between \SIrange{8}{12}{\micro\meter}, but its exact shape and location are dependent on the mineralogy of the rock and the constituent minerals' silicate polymerization (Figure~\ref{fig:all emi comp plots}). The mafic powders shown in Figure~\ref{fig:compositional variety} each have a different Si-O feature shape and center location. The differences in shape and center location between the three stems from differences in their mineral breakdowns, such as their plagioclase and pyroxene content. 

The detectability of this feature in exoplanet emission spectra is strongly dependent on its depth, which varies significantly across our sample set. One of the most important determining factors for the depth of this feature is the size of individual mineral grains in the sample (note that this is distinct from the sample texture, which is a macroscopic material property). For igneous rocks, the grain sizes reflect the cooling history of the sample; materials that cool faster will typically have smaller grain sizes, while those that cool more slowly will have larger grain sizes. To illustrate this effect, we measured the depth of the Si-O bending feature in model exoplanet emission spectra for each slab surface type, using the same LHS 3844~b system parameters as before. For the dunite xenolith we modeled the crushed texture, as this sample was not available in slab form.  We calculated the depth of this feature by comparing the wavelength-dependent planet-star flux ratio for each surface type with a blackbody emission model at the same temperature (\SI{1000}{\kelvin}), and recorded the maximum depth of the Si-O bending feature shortward of \SI{12.5}{\micro\meter}. 

We find that the crushed dunite xenolith and the orlando gold granite slab have significantly deeper Si-O stretching features ($>150$ ppm decrease in planet-star flux ratio in the deepest part of the feature) than any of our other samples. This is in good agreement with our expectations based on visual inspection, as both samples have grains large enough to see with the naked eye. The basaltic andesite, K1919 basalt, and andesite (STM-101) have the smallest feature depths ($<70$ ppm decrease in planet-star flux ratio in the deepest part of the feature); these are all very fine-grained rocks with no visible crystals.


Next, we quantified the detectability of the Si-O stretching feature in \textit{JWST} observations, using the same methodology as in \S\ref{sec:olivine feature} and \S\ref{sec:transparency feature}. We selected the dalmatian granite sample for this test, as it displays one of the deepest Si-O stretching features in our spectroscopic library. We utilized the slab texture in order to eliminate the transparency feature, which partially overlaps with this feature. We then calculated the number of eclipses necessary to detect the Si-O feature (see Fig~\ref{fig:textures} for a plot of the modeled emission spectrum corresponding to this surface). We specified a wavelength resolution of $R=10$, and only used data up to \SI{12.5}{\micro\meter} for these simulations, as the predicted measurement errors become so large at longer wavelengths that those points contribute negligibly to the fit. For this test, the null hypothesis was a blackbody surface with the temperature as a free parameter. We calculated the detection significance for the feature by adding a second free parameter that scales the amplitude of the Si-O feature within its respective wavelengths. We then carried out the nested sampling fit following the same procedure as in \S\ref{sec:olivine feature}. We repeated these fits on 1000 different iterations of the \texttt{Pandexo} generated data, and found that at least five eclipse observations were necessary to detect the Si-O feature to $\geq3\sigma$ confidence level. This is more favorable than any of the other features considered, albeit with the caveat that most other samples have shallower Si-O streching features than the dalmatian granite. Again, we also performed this same test on the EG-19-63 olivine clinopyroxenite, STM-101 andesite, and K1919 basalt slabs to understand the range in detectability of the feature. The different surface types required 8, 12, and $>$30 eclipses, respectively. This suggests that the Si-O stretching feature is only likely to be detectable for the subset of surface types with moderate to large mineral grain sizes. On the other hand, \citet{First2024} demonstrated that it is possible to detect the aqueous alteration features present in their basaltic samples with the same number of eclipses we found for detecting the Si-O feature for the dalmatian granite of five eclipses. 

\subsubsection{Christiansen Feature}
\label{sec:christiansen feature}
The Christiansen Feature (CF) corresponds to the wavelength region where the refractive index of the material approaches the refractive index of its surrounding medium (e.g., air or vacuum), resulting in a reflectance minimum (emissivity maximum) between \SIrange{7}{9}{\micro\meter}. This feature marks the beginning of the Si-O stretching feature discussed in \S\ref{sec:SiO stretching feature}. The location of this feature in wavelength space varies as a function of the polymerization of silicates (number of shared oxygen in \ce{SiO4} tetrahedra in the mineral structure), which coarsely tracks to \ce{SiO} content of the material \citep{Conel1969}, making it a useful diagnostic for determining the composition of bare-rock surfaces. Materials with a higher \ce{SiO2} wt\% will have a CF located closer to \SI{7}{\micro\meter}, while materials with a lower \ce{SiO2} wt\% will have a CF located closer to \SI{9}{\micro\meter}. The CF is frequently used as a diagnostic for surface composition in studies of solar system bodies \citep[e.g.,][]{Glotch2010}, and has played an important role in the design of several missions including the Lunar Reconnaissance Orbiter \citep{Paige2010} and the BepiColombo mission to Mercury \citep{Hiesinger2020} because even a few judiciously-placed spectral channels can discriminate ultramafic, mafic, feldspathic, and silic lithologies. However, \citet{Fortin2024} found that there was not a clear shift in the location of the CF with the \ce{SiO2} content alone for their samples when measured at high temperatures, so they derived an empirical model between the CF and the composition of the rock including its \ce{Al2O3}, \ce{FeO}, and \ce{CaO} contents in addition to the \ce{SiO2} content. This model is generally consistent with their samples, but the measured CFs are still quite scattered around the linear fit.

\begin{figure*}[ht!]
    \centering
    \includegraphics[width=\textwidth]{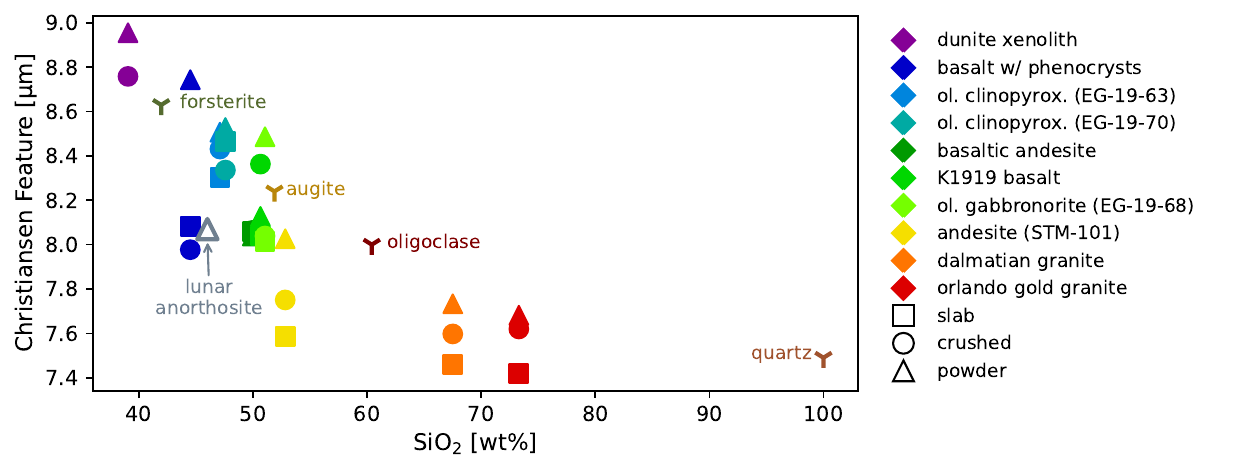}
    \caption{The location of the Christiansen Feature as a function of the \ce{SiO2} wt\% for the silicate samples in our spectral library. The textures (slab, crushed, and powder) of each sample are denoted by the marker shapes (square, circle, and triangle; respectively). We also include labeled markers showing the CF of four pure minerals for comparison: forsterite (olivine), augite (pyroxene), oligoclase (plagioclase), and quartz. Since we do not have a feldspathic sample in our library, we include the CF of the lunar anorthosite powder from \citet{Hu2012} as an unfilled triangle in grey.}
    \label{fig:CF}
\end{figure*}

We use the information from our compositional analysis to plot the location of the CF in our hemispherical reflectance measurements as a function of their \ce{SiO2} wt\% in Figure~\ref{fig:CF}. We also overplot the CF locations for four pure minerals, including forsterite, augite, oligoclase (a type of plagioclase), and quartz, and the lunar anorthosite powder from \citet{Hu2012} in the figure for comparison. We obtained the thermal emission spectra for the four minerals from the ASU Thermal Emission Spectroscopy Laboratory Spectral Library (sample identifications BUR-3720A, HS-119.4B, BUR-060, and Quartz, respectively). The forsterite and oligoclase measurements were for powders with \SIrange{710}{1000}{\micro\meter} grains, the augite measurements was for a powder with \SIrange{250}{1000}{\micro\meter}, and the quartz measurement was for a powder with \SI{125}{\micro\meter} to \SI{2}{\milli\meter} grains.

As expected, we see a quasi-linear relationship between the CF wavelength location and the \ce{SiO2} wt\% of each sample. The pure mineral samples show a broadly similar trend, but do not exactly match the trend for our samples. This is not surprising, as each sample contains a different mixture of \ce{SiO2}-bearing minerals, each with their own distinct CF locations. 
The texture of the material also affects the CF location, as demonstrated by published studies of pure minerals \citep{Logan1973, Mustard1997} and rocks \citep{Cooper2002}. For rocks, the authors found that the CF shifts to longer wavelengths for powdered material relative to solid material, exactly matching the behavior that we see in our measurements. This is because the Si-O stretching feature with the CF at the shortest wavelength will dominate the spectra of solid material, while the volume-scattering increase and Si-O feature contrast decrease of powdered material causes the CFs at longer wavelengths to dominate instead \citep{Cooper2002}. 

\begin{figure}[ht!]
    \centering
    \includegraphics[width = 0.47\textwidth]{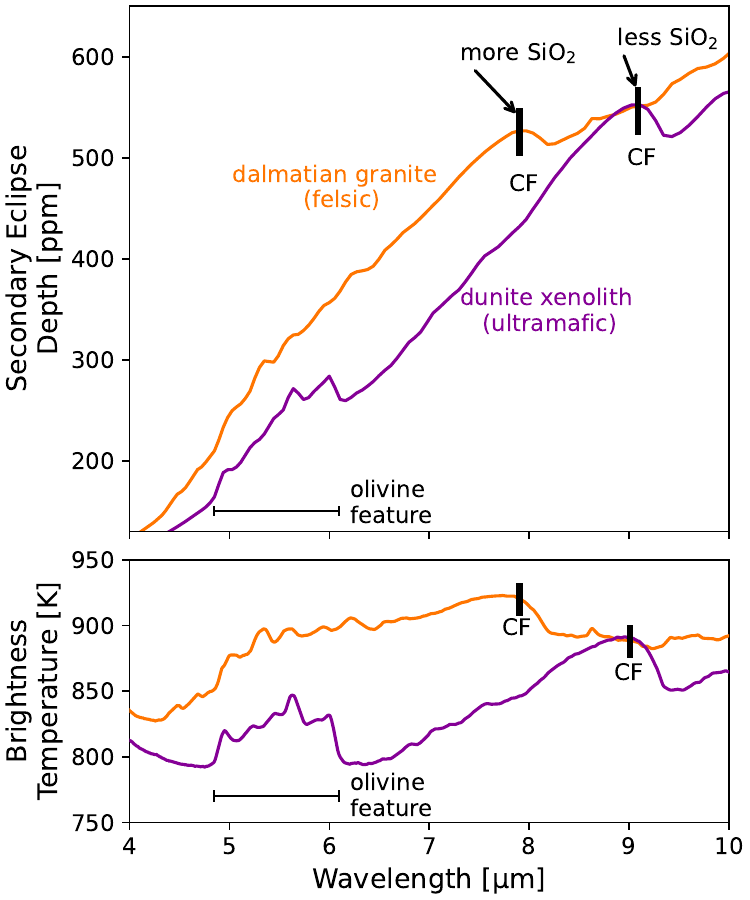}
    \caption{Predicted wavelength-dependent secondary eclipse depths for a LHS 3844 b analog assuming a blackbody for the stellar spectrum. We show two models spanning the range from lowest (dunite xenolith) to the second highest (dalmatian granite) \ce{SiO2} wt\%. We modeled the dalmatian granite instead of the orlando gold granite (the sample with the highest \ce{SiO2} wt\%) due to its lower reflectivity allowing for an easier comparison with the dunite xenolith sample. The changing location of the CF in these two models reflects their differing \ce{SiO2} contents.}
    \label{fig:CF depths}
\end{figure}

We illustrate the changing location of this feature for a LHS 3844~b analog in Figure~\ref{fig:CF depths}, focusing on dalmatian granite and the dunite xenolith as the two samples with the most widely varying \ce{SiO2} wt\%. 
We quantified our ability to measure the changing location of this feature by adding an additional free parameter to our parametric Si-O feature model from \S\ref{sec:SiO stretching feature} that allows the location of the Si-O feature to shift in wavelength space. We simulated \textit{JWST} observations of the dalmatian granite, STM-101 andesite, and EG-19-63 olivine clinopyroxenite slab surface models, and set the number of eclipses to the value required to achieve a $3\sigma$ detection of the Si-O feature (5, 12, and 8, respectively). We did not model the K1919 basalt surface, as its Si-O feature would require more than 30 eclipses to achieve a statistically significant detection. We repeated the same nested sampling routine on 1000 different iterations of each surface type, and found that the CF can be constrained with $1\sigma$ uncertainties of \SI{0.3}{\micro\meter} (dalmatian granite), \SI{0.7}{\micro\meter} (STM-101 andesite), and \SI{0.5}{\micro\meter} (EG-19-63 olivine clinopyroxenite). These uncertainties are small enough to allow us to roughly differentiate between ultramafic, mafic, intermediate, and felsic igneous rock types, but more observations will allow us to pin down their wavelength locations more precisely.

\subsection{Temperature-Dependent Changes in Spectral Shape}\label{sec:highT_discussion}
\begin{figure}[ht!]
    \centering
    \includegraphics[width=0.47\textwidth]{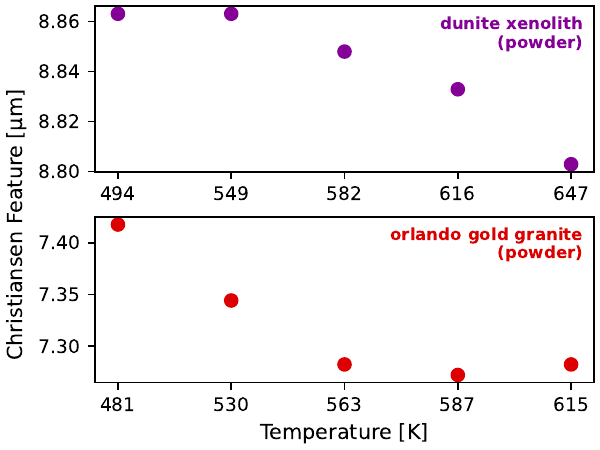}
    \caption{Measured Christiansen Feature location as a function of temperature for the dunite xenolith powder (top) and the orlando gold granite powder (bottom). Note the different horizontal axis ranges.}
    \label{fig:CF shift}
\end{figure}

When we heat the dunite xenolith powder, we see that the measured depth of the Si-O stretching feature increases smoothly with increasing temperature, while the depth of the transparency feature (due to multiple scattering between grains) is constant across all temperatures. We see a similar behavior in the orlando gold granite powder, which also exhibits a clear temperature dependence in the measured depth of the Si-O stretching feature. This behavior is consistent with the spectra in \citet{Fortin2024} where the Si-O feature of all of their samples becomes deeper as a function of temperature. K1919 basalt powder does not show an obvious temperature dependence in the depth of its Si-O stretching feature, but this feature is also much shallower than in the other two powders and it is therefore likely that any temperature-dependent changes lie below our detection threshold. 
For the dunite xenolith and orlando gold granite powders, we also observe a shift in the wavelength location of the CF (see \S\ref{sec:christiansen feature} for a detailed definition) as a function of temperature. In both samples the CF shifts to shorter wavelengths as temperature increases (see Figure~\ref{fig:CF shift}), in good agreement with results from previous studies \citep[e.g.,][]{DonaldsonHanna2017, Ferrari2020}.

We do not see temperature-dependent trends in the depths or locations of spectral features in any of the other samples, including the orlando gold granite slab and the crushed dunite xenolith. While it is possible that the crushed dunite xenolith suffered from additional flux contamination from the cup as a result of its relatively high transparency, this is not the case for the orlando gold granite slab. We speculate that thermophyscial properties of powders may enhance the amplitude of temperature dependent changes. Regardless of its origin, the fact that we see little to no change in feature depth for most of our samples, and that the changes in feature depth for the dunite xenolith and orlando gold granite powders are relatively small, suggests that 
these effects likely lie beneath the noise floor of \textit{JWST} observations of hot, rocky exoplanet surfaces.

This statement comes with one significant caveat. Most exoplanet emission models, including our own, are constructed by converting room-temperature reflectance measurements into emissivity measurements.  When we compare the feature depths in emissivities calculated using this method versus those measured directly in emission, we 
see a systematic discrepancy in that the spectral constrast is lower in our emissivity data. We note that for our more heterogeneous slab samples, it is possible that we measured a slightly different location on the sample in reflectance versus emission. Similarly, particulate samples may have had slightly different packing (porosity) between the two measurements. Both of these effects can lead to differences in the measured spectral properties. However, neither effect can explain why our reflectance derived emissivities consistently display deeper features than the measured high temperature emissivities, regardless of sample texture or (for slabs) heterogeneity.

Previous studies that carried out equivalent comparisons have also noted that absorption bands measured in reflectance display deeper contrasts than those measured in emissivity \citep[e.g.,][]{Salisbury1994, Hapke2012, Ferrari2020}. \citet{Hapke2012} enumerated several potential explanations, including the possibility that increased multiple scattering in emissivity measurements might cause the depth of bands to diminish. More generally, Kirchhoff's law of thermal radiation tells us that $1-R = \varepsilon$ when the system is in thermal equilibrium \citep{Hapke2012}. Our emissivity measurements violate this assumption, because the surrounding background is not at the same temperature as the sample. Furthermore, our method of heating the sample via induction from beneath likely resulted in an internal temperature gradient within the sample. This effect was previously explored in \citet{DonaldsonHanna2017}, which performed controlled emissivity measurements to reproduce the spectra of rocky surfaces on the Moon, including the effect of a temperature gradient in the upper surface layer. Controlling for atmospheric pressure alone, the authors observed a change in the contrast between the CF and Si-O stretching regions.

We conclude that the depths of features measured in emission are dependent on exact experimental set-up, including the atmospheric pressure, the temperature of the sample, the temperature of the surrounding medium, and how the sample is heated. It is likely that some or all of these effects may contribute to the differences between feature depths measured in reflection versus emission for our samples. However, none of these effects alter the fundamental nature of these features, and their presence or absence in the emission spectra of rocky exoplanets is therefore a reliable diagnostic of the surface properties of these objects with the caveat that shallower features will be more difficult to robustly identify in observations with substantial noise.

\subsection{LHS 3844 b as a Case Study}\label{sec:LHS3844b}
\begin{figure} 
    \centering
    \includegraphics[width=0.47\textwidth]{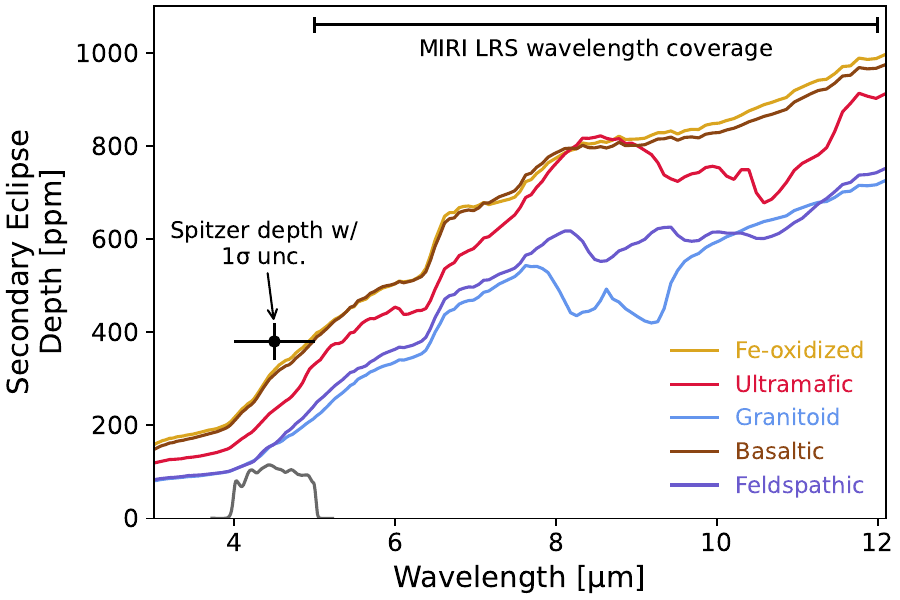}
    \caption{Predicted wavelength-dependent secondary eclipse depths for the super-Earth LHS~3844~b generated using the powdered samples from \citet{Hu2012} and an updated stellar spectrum.  As discussed in \S\ref{sec:sample_selection}, we only show the subset of surfaces relevant for hot, rocky exoplanets.
    We overplot the published \textit{Spitzer} measurement from \citet{Kreidberg2019} as a black circle with the corresponding $1\sigma$ uncertainty shown as a vertical error bar.  The horizontal error bar indicates the approximate wavelength range of the \SI{4.5}{\micro\meter} \textit{Spitzer} bandpass, and the grey curve corresponds to the instrument's spectral response function.}
    \label{fig:HES2012_surfaces}
\end{figure}

As introduced in \S\ref{sec:intro}, LHS 3844 b is one of the most favorable rocky exoplanet targets for emission spectroscopy with \textit{JWST}. Published \SI{4.5}{\micro\meter} photometric phase curve measurements with \textit{Spitzer} revealed that the planet has little to no atmosphere, and \citet{Kreidberg2019} found that its measured dayside brightness temperature was most consistent with a basaltic surface when comparing to the spectral library from \citet{Hu2012}. We generate new models for this planet using an updated SPHINX model with an effective temperature of 3000~K, log($g$ [\unit{\centi\meter\per\second\squared}]) of 5, and solar C/O and metallicity for the host star, which provides a good match to the stellar spectrum measured with MIRI LRS (Zieba et al. in prep.). We begin by modeling the same set of surfaces presented in \citet{Hu2012}, in order to quantify the impact of this change on our interpretation of the Spitzer photometry and demonstrate the degeneracies between albedo and surface properties. The resulting emission spectra are shown in  Figure~\ref{fig:HES2012_surfaces}. We find that the new basaltic surface model still provides the best match to this measurement, but now only agrees to within $<2\sigma$ rather than the previous $<1\sigma$. \citet{First2024} also generated a similar set of models for their basaltic samples, and found that they were all consistent with the previous \textit{Spitzer} measurement.

\begin{figure*}[ht!]
    \centering
    \includegraphics[width=\textwidth]{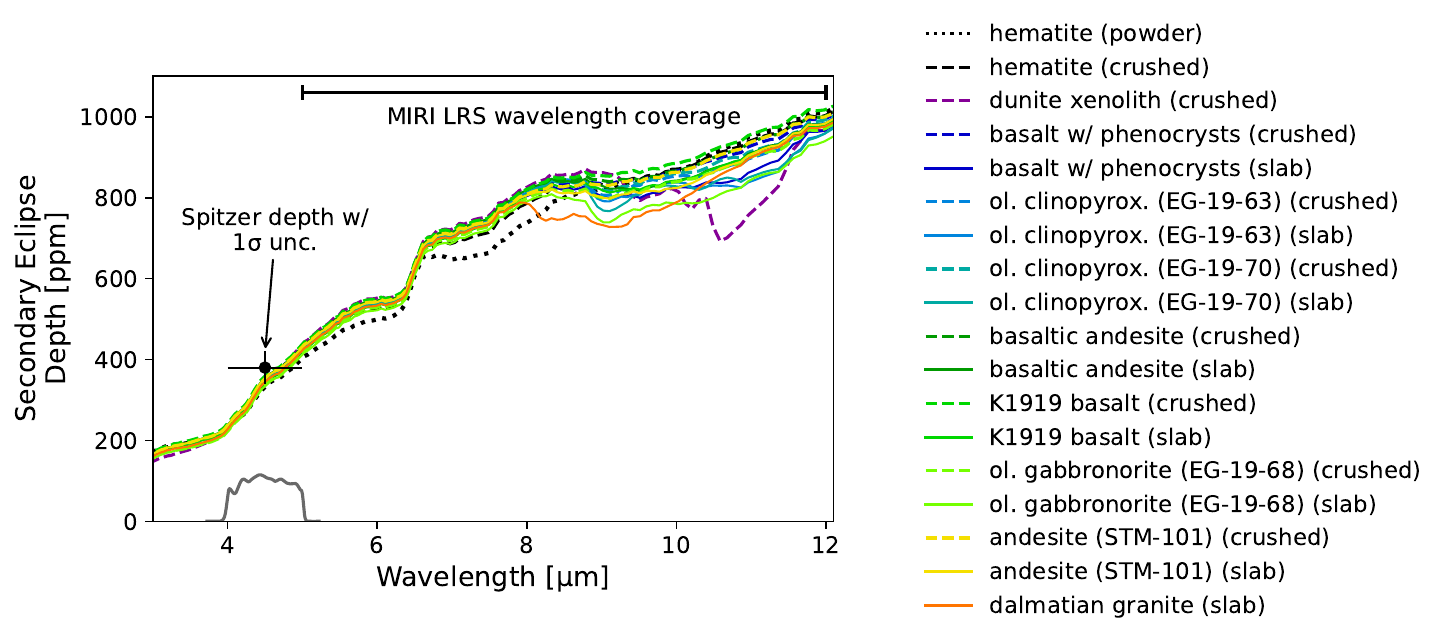}
    \caption{Thermal emission spectra of 18/31 surface and texture combinations in our spectral library that are $\leq2\sigma$ consistent with the \SI{4.5}{\micro\meter} \textit{Spitzer} measurement (the marker in black) for LHS 3844 b in \citet{Kreidberg2019}. These models are generated for the LHS 3844 b system with a SPHINX stellar model.}
    \label{fig:consistent_surfaces}
\end{figure*}

Next, we explore whether our updated spectral library can provide an improved match to this measurement. With our 11 samples and variety of textures, there are 31 possible combinations. Of those, 18 are $\leq2\sigma$ consistent with the \textit{Spitzer} measurement in Figure~\ref{fig:consistent_surfaces}. These include the hematite, dunite xenolith, basalt with phenocrysts, olivine clinopyroxenite (EG-19-63), olivine clinopyroxenite (EG-19-70), basaltic andesite, K1919 basalt, olivine gabbronorite (EG-19-68), andesite (STM-101), and the dalmatian granite. For all but one surface type, the data prefer the slab and/or crushed textures. The hematite is the only sample for which the powdered texture is consistent with the data. As expected, we prefer the subset of models that have the lowest wavelength-averaged optical and near-infrared albedos and the correspondingly highest surface temperatures. 

With our expanded spectral library, there are now representative samples from multiple compositional classes that are at least $2\sigma$ consistent with the \textit{Spitzer} measurement. This is weaker than the conclusions in \cite{Kreidberg2019}, which used the spectral library from \cite{Hu2012} to argue that feldspathic and granitoid materials provided a relatively poor fit to this measurement. We conclude that when we allow for a greater diversity of compositions and textures, a single broadband measurement is of limited use for constraining the surface compositions of rocky exoplanets.


\section{Conclusions}
In this work we present a new spectral library for modeling the thermal emission spectra of hot, rocky exoplanets with little to no atmosphere. We collect new reflectance and emissivity measurements for a set of igneous rock samples spanning a wide range in SiO$_2$ abundances, including samples in all four compositional classes (ultramafic, mafic, intermediate, and felsic), as well as one Fe-oxidized sample.  
We measure the reflectances of these samples in three textures, including solid slabs, coarsely crushed material (\SI{500}{\micro\meter} to \SI{1}{\milli\meter}), and finely ground powder material (\SI{25}{\micro\meter} to \SI{63}{\micro\meter}). We also directly measure the mid-infrared emissivities of a subset of ten samples at temperatures ranging between $500-800$~K, analogous to the temperatures of many rocky exoplanets. 

We incorporate both the old spectral library from \citet{Hu2012} and our new spectral library into an updated version of the open-source python package \texttt{PLATON} \citep{Zhang2019, Zhang2020} in order to calculate the thermal emission spectra of close-in rocky exoplanets. Our model also accounts for contributions from reflected light, which are typically only important shortward of \SI{5}{\micro\meter}. This new version of the \texttt{PLATON} package is be  publicly available in version 6.3.  

We use our new spectral library and modeling framework to demonstrate that samples with similar chemical compositions can have significantly different reflectance spectra, with correspondingly large differences in dayside albedos and surface temperatures. 
We also quantify how changes in texture can alter the albedos of our samples. Because powdered materials can be significantly more reflective at optical and near-infrared wavelengths than slabs of the same material, these large differences in reflectance directly translate to a wide range of predicted dayside temperatures.  We conclude that albedo is a relatively weak and degenerate indicator of surface properties. 

We identify a select set of the most promising spectral features for constraining the surface composition and texture of rocky exoplanets. We focus on features located in the mid-infrared, as the higher planet-star flux ratio at these wavelengths makes them easier to detect than the near-infrared spectral features commonly used in studies of solar system objects. We identify four promising features for future mid-infrared spectroscopic studies, including the \SI{5.6}{\micro\meter} olivine feature, the transparency feature (a key diagnostic of grain size), the Si-O stretching feature (the location of this feature changes as a function of \ce{SiO2} wt\%), and the Christiansen feature. 
Finally, we use our high-temperature emissivity measurements to demonstrate that the spectral shapes of features at mid-infrared wavelengths are only weakly dependent on temperature. 
However, we also find that the depths of features measured in emission are systematically shallower than expected based on their room-temperature reflectance measurements.  This is likely due to a combination of factors, including increased multiple scattering in the emission spectra, the degree of vacuum, and the presence of internal temperature gradients within our samples.

Finally, we provide an updated interpretation of published \SI{4.5}{\micro\meter} photometry of dayside emission from the benchmark super-Earth LHS 3844~b from \cite{Kreidberg2019}. After incorporating an updated spectrum for the host star, we find that this planet's dayside emission can be matched by most rocks and coarse-grained materials in our spectral library but cannot be matched by fine-grained silicate powders. The compositions that match the observations range from granitic to ultramafic, including the \citet{Hu2012} surface types, both silicates and oxidized iron. This is contrary to the interpretation published by \citet{Kreidberg2019}, who concluded that feldspathic and granitoid surfaces were a poor match ($\geq3\sigma$ different) to the data. This highlights the importance of texture as well as composition when modeling the fluxes expected from different types.

Our new modeling framework and spectral library provides important new capabilities to model the bare-rock spectra of  \textit{JWST} targets, and be used to inform observing strategies for future spectroscopic observing programs with \textit{JWST}. These observations offer an unprecedented opportunity to obtain improved constraints on the surface compositions of rocky exoplanets, providing a new window into the geological evolution of exoplanets.

\section*{acknowledgments}
This research was carried out at the Jet Propulsion Laboratory and the California Institute of Technology under a contract with the National Aeronautics and Space Administration and funded through the President’s and Director’s Research \& Development Fund Program. HAK and KP also gratefully acknowledge funding support from the Caltech Center for Comparative Planetary Evolution. KP thanks Paul Asimow, Claire Bucholz, Ken Farley, Shane Houchin, William Lawrence, Juliet Ryan-Davis, Ed Stolper, and Oliver Wilner for shaping our rock library by providing their geological expertise and/or samples, and Nick Anderson, Thomas Bailey, Rachael Danyew, Mark Garcia, Rebecca Greenberger, Martin Mendez, Zachariah Milby, Yuyu Phua, and Morgan Saidel for lending a hand with sample preparation and/or shipment.
\vspace{5mm}

\bibliography{sample631}{}

\begin{thebibliography}{}
\expandafter\ifx\csname natexlab\endcsname\relax\def\natexlab#1{#1}\fi
\providecommand{\url}[1]{\href{#1}{#1}}
\providecommand{\dodoi}[1]{doi:~\href{http://doi.org/#1}{\nolinkurl{#1}}}
\providecommand{\doeprint}[1]{\href{http://ascl.net/#1}{\nolinkurl{http://ascl.net/#1}}}
\providecommand{\doarXiv}[1]{\href{https://arxiv.org/abs/#1}{\nolinkurl{https://arxiv.org/abs/#1}}}

\bibitem[{{Anderson} {et~al.}(2017){Anderson}, {Ehlmann}, {Forni}, {Clegg}, {Cousin}, {Thomas}, {Lasue}, {Delapp}, {McInroy}, {Gasnault}, {Dyar}, {Schr{\"o}der}, {Maurice}, \& {Wiens}}]{Anderson2017}
{Anderson}, D.~E., {Ehlmann}, B.~L., {Forni}, O., {et~al.} 2017, Journal of Geophysical Research (Planets), 122, 744, \dodoi{10.1002/2016JE005164}

\bibitem[{{Batalha} {et~al.}(2017){Batalha}, {Mandell}, {Pontoppidan}, {Stevenson}, {Lewis}, {Kalirai}, {Earl}, {Greene}, {Albert}, \& {Nielsen}}]{Batalha2017}
{Batalha}, N.~E., {Mandell}, A., {Pontoppidan}, K., {et~al.} 2017, \pasp, 129, 064501, \dodoi{10.1088/1538-3873/aa65b0}

\bibitem[{{Bowey} \& {Hofmeister}(2005)}]{Bowey2005}
{Bowey}, J.~E., \& {Hofmeister}, A.~M. 2005, \mnras, 358, 1383, \dodoi{10.1111/j.1365-2966.2005.08848.x}

\bibitem[{{Brinkman} {et~al.}(2024{\natexlab{a}}){Brinkman}, {Polanski}, {Huber}, {Weiss}, {Valencia}, \& {Plotnykov}}]{Brinkman2024a}
{Brinkman}, C.~L., {Polanski}, A.~S., {Huber}, D., {et~al.} 2024{\natexlab{a}}, arXiv e-prints, arXiv:2409.08361, \dodoi{10.48550/arXiv.2409.08361}

\bibitem[{{Brinkman} {et~al.}(2024{\natexlab{b}}){Brinkman}, {Weiss}, {Huber}, {Lee}, {Kolecki}, {Tenn}, {Zhang}, {Narayanan}, {Polanski}, {Dai}, {Bean}, {Beard}, {Brady}, {Brodheim}, {Brown}, {Deich}, {Edelstein}, {Fulton}, {Giacalone}, {Gibson}, {Gilbert}, {Halverson}, {Handley}, {Hill}, {Holcomb}, {Holden}, {Householder}, {Howard}, {Isaacson}, {Kaye}, {Laher}, {Lanclos}, {Ong}, {Payne}, {Petigura}, {Pidhorodetska}, {Poppett}, {Roy}, {Rubenzahl}, {Saunders}, {Schwab}, {Seifahrt}, {Shaum}, {Sirk}, {Smith}, {Smith}, {Stef{\'a}nsson}, {St{\"u}rmer}, {Thorne}, {Turtelboom}, {Tyler}, {Valliant}, {Van Zandt}, {Walawender}, {Yee}, {Yeh}, \& {Zink}}]{Brinkman2024b}
{Brinkman}, C.~L., {Weiss}, L.~M., {Huber}, D., {et~al.} 2024{\natexlab{b}}, arXiv e-prints, arXiv:2410.00213, \dodoi{10.48550/arXiv.2410.00213}

\bibitem[{{Charlier} {et~al.}(2013){Charlier}, {Grove}, \& {Zuber}}]{Charlier2013}
{Charlier}, B., {Grove}, T.~L., \& {Zuber}, M.~T. 2013, Earth and Planetary Science Letters, 363, 50, \dodoi{10.1016/j.epsl.2012.12.021}

\bibitem[{{Christensen} {et~al.}(2001){Christensen}, {Bandfield}, {Hamilton}, {Ruff}, {Kieffer}, {Titus}, {Malin}, {Morris}, {Lane}, {Clark}, {Jakosky}, {Mellon}, {Pearl}, {Conrath}, {Smith}, {Clancy}, {Kuzmin}, {Roush}, {Mehall}, {Gorelick}, {Bender}, {Murray}, {Dason}, {Greene}, {Silverman}, \& {Greenfield}}]{Christensen2001}
{Christensen}, P.~R., {Bandfield}, J.~L., {Hamilton}, V.~E., {et~al.} 2001, \jgr, 106, 23823, \dodoi{10.1029/2000JE001370}

\bibitem[{Clark(1999)}]{Clark1999}
Clark, R.~N. 1999, Spectroscopy of Rocks and Minerals and Principles of Spectroscopy (Wiley), 3–58

\bibitem[{{Conel}(1969)}]{Conel1969}
{Conel}, J.~E. 1969, \jgr, 74, 1614, \dodoi{10.1029/JB074i006p01614}

\bibitem[{{Cooper} {et~al.}(2002){Cooper}, {Salisbury}, {Killen}, \& {Potter}}]{Cooper2002}
{Cooper}, B.~L., {Salisbury}, J.~W., {Killen}, R.~M., \& {Potter}, A.~E. 2002, Journal of Geophysical Research (Planets), 107, 5017, \dodoi{10.1029/2000JE001462}

\bibitem[{{Crossfield} {et~al.}(2022){Crossfield}, {Malik}, {Hill}, {Kane}, {Foley}, {Polanski}, {Coria}, {Brande}, {Zhang}, {Wienke}, {Kreidberg}, {Cowan}, {Dragomir}, {Gorjian}, {Mikal-Evans}, {Benneke}, {Christiansen}, {Deming}, \& {Morales}}]{Crossfield2022}
{Crossfield}, I. J.~M., {Malik}, M., {Hill}, M.~L., {et~al.} 2022, \apjl, 937, L17, \dodoi{10.3847/2041-8213/ac886b}

\bibitem[{{Dai} {et~al.}(2019){Dai}, {Masuda}, {Winn}, \& {Zeng}}]{Dai2019}
{Dai}, F., {Masuda}, K., {Winn}, J.~N., \& {Zeng}, L. 2019, \apj, 883, 79, \dodoi{10.3847/1538-4357/ab3a3b}

\bibitem[{{Domingue} {et~al.}(2014){Domingue}, {Chapman}, {Killen}, {Zurbuchen}, {Gilbert}, {Sarantos}, {Benna}, {Slavin}, {Schriver}, {Tr{\'a}vn{\'\i}{\v{c}}ek}, {Orlando}, {Sprague}, {Blewett}, {Gillis-Davis}, {Feldman}, {Lawrence}, {Ho}, {Ebel}, {Nittler}, {Vilas}, {Pieters}, {Solomon}, {Johnson}, {Winslow}, {Helbert}, {Peplowski}, {Weider}, {Mouawad}, {Izenberg}, \& {McClintock}}]{Domingue2014}
{Domingue}, D.~L., {Chapman}, C.~R., {Killen}, R.~M., {et~al.} 2014, \ssr, 181, 121, \dodoi{10.1007/s11214-014-0039-5}

\bibitem[{{Donaldson Hanna} {et~al.}(2017){Donaldson Hanna}, {Greenhagen}, {Patterson}, {Pieters}, {Mustard}, {Bowles}, {Paige}, {Glotch}, \& {Thompson}}]{DonaldsonHanna2017}
{Donaldson Hanna}, K.~L., {Greenhagen}, B.~T., {Patterson}, W.~R., {et~al.} 2017, \icarus, 283, 326, \dodoi{10.1016/j.icarus.2016.05.034}

\bibitem[{Ehlmann \& Edwards(2014)}]{Ehlmann2014}
Ehlmann, B.~L., \& Edwards, C.~S. 2014, Annual Review of Earth and Planetary Sciences, 42, 291, \dodoi{https://doi.org/10.1146/annurev-earth-060313-055024}

\bibitem[{{Elkins-Tanton} {et~al.}(2005){Elkins-Tanton}, {Hess}, \& {Parmentier}}]{Elkins-Tanton2005}
{Elkins-Tanton}, L.~T., {Hess}, P.~C., \& {Parmentier}, E.~M. 2005, Journal of Geophysical Research (Planets), 110, E12S01, \dodoi{10.1029/2005JE002480}

\bibitem[{{Ferrari} {et~al.}(2020){Ferrari}, {Maturilli}, {Carli}, {D'Amore}, {Helbert}, {Nestola}, \& {Hiesinger}}]{Ferrari2020}
{Ferrari}, S., {Maturilli}, A., {Carli}, C., {et~al.} 2020, Earth and Planetary Science Letters, 534, 116089, \dodoi{10.1016/j.epsl.2020.116089}

\bibitem[{{First} {et~al.}(2024){First}, {Mishra}, {Gazel}, {Lewis}, {Letai}, \& {Hanssen}}]{First2024}
{First}, E.~C., {Mishra}, I., {Gazel}, E., {et~al.} 2024, Nature Astronomy, \dodoi{10.1038/s41550-024-02412-7}

\bibitem[{{Fortin} {et~al.}(2024){Fortin}, {Gazel}, {Williams}, {Thompson}, {Kaltenegger}, \& {Ramsey}}]{Fortin2024}
{Fortin}, M.-A., {Gazel}, E., {Williams}, D.~B., {et~al.} 2024, \apjl, 974, L7, \dodoi{10.3847/2041-8213/ad7d89}

\bibitem[{{Gaillard} {et~al.}(2022){Gaillard}, {Bernadou}, {Roskosz}, {Bouhifd}, {Marrocchi}, {Iacono-Marziano}, {Moreira}, {Scaillet}, \& {Rogerie}}]{Gaillard2022}
{Gaillard}, F., {Bernadou}, F., {Roskosz}, M., {et~al.} 2022, Earth and Planetary Science Letters, 577, 117255, \dodoi{10.1016/j.epsl.2021.117255}

\bibitem[{{Glotch} {et~al.}(2010){Glotch}, {Lucey}, {Bandfield}, {Greenhagen}, {Thomas}, {Elphic}, {Bowles}, {Wyatt}, {Allen}, {Hanna}, \& {Paige}}]{Glotch2010}
{Glotch}, T.~D., {Lucey}, P.~G., {Bandfield}, J.~L., {et~al.} 2010, Science, 329, 1510, \dodoi{10.1126/science.1192148}

\bibitem[{{Greene} {et~al.}(2023){Greene}, {Bell}, {Ducrot}, {Dyrek}, {Lagage}, \& {Fortney}}]{Greene2023}
{Greene}, T.~P., {Bell}, T.~J., {Ducrot}, E., {et~al.} 2023, \nat, 618, 39, \dodoi{10.1038/s41586-023-05951-7}

\bibitem[{{Gressier} {et~al.}(2024){Gressier}, {Espinoza}, {Allen}, {Sing}, {Banerjee}, {Barstow}, {Valenti}, {Lewis}, {Birkmann}, {Challener}, {Manjavacas}, {Alves de Oliveira}, {Crouzet}, \& {Beck}}]{Gressier2024}
{Gressier}, A., {Espinoza}, N., {Allen}, N.~H., {et~al.} 2024, \apjl, 975, L10, \dodoi{10.3847/2041-8213/ad73d1}

\bibitem[{{Hammond} {et~al.}(2024){Hammond}, {Guimond}, {Lichtenberg}, {Nicholls}, {Fisher}, {Luque}, {Meier}, {Taylor}, {Changeat}, {Dang}, {Herbort}, \& {Teske}}]{Hammond2024}
{Hammond}, M., {Guimond}, C.~M., {Lichtenberg}, T., {et~al.} 2024, arXiv e-prints, arXiv:2409.04386, \dodoi{10.48550/arXiv.2409.04386}

\bibitem[{{Hansen}(2008)}]{Hansen2008}
{Hansen}, B. M.~S. 2008, \apjs, 179, 484, \dodoi{10.1086/591964}

\bibitem[{{Hapke}(2001)}]{Hapke2001}
{Hapke}, B. 2001, \jgr, 106, 10039, \dodoi{10.1029/2000JE001338}

\bibitem[{{Hapke}(2012)}]{Hapke2012}
---. 2012, {Theory of Reflectance and Emittance Spectroscopy} (Cambridge University Press), \dodoi{10.1017/CBO9781139025683}

\bibitem[{{Helbert} {et~al.}(2013){Helbert}, {Nestola}, {Ferrari}, {Maturilli}, {Massironi}, {Redhammer}, {Capria}, {Carli}, {Capaccioni}, \& {Bruno}}]{Helbert2013}
{Helbert}, J., {Nestola}, F., {Ferrari}, S., {et~al.} 2013, Earth and Planetary Science Letters, 371, 252, \dodoi{10.1016/j.epsl.2013.03.038}

\bibitem[{{Hiesinger} {et~al.}(2020){Hiesinger}, {Helbert}, {Alemanno}, {Bauch}, {D'Amore}, {Maturilli}, {Morlok}, {Reitze}, {Stangarone}, {Stojic}, {Varatharajan}, {Weber}, \& {Mertis Co-I Team}}]{Hiesinger2020}
{Hiesinger}, H., {Helbert}, J., {Alemanno}, G., {et~al.} 2020, \ssr, 216, 110, \dodoi{10.1007/s11214-020-00732-4}

\bibitem[{{Hu} {et~al.}(2012){Hu}, {Ehlmann}, \& {Seager}}]{Hu2012}
{Hu}, R., {Ehlmann}, B.~L., \& {Seager}, S. 2012, \apj, 752, 7, \dodoi{10.1088/0004-637X/752/1/7}

\bibitem[{{Hu} {et~al.}(2024){Hu}, {Bello-Arufe}, {Zhang}, {Paragas}, {Zilinskas}, {van Buchem}, {Bess}, {Patel}, {Ito}, {Damiano}, {Scheucher}, {Oza}, {Knutson}, {Miguel}, {Dragomir}, {Brandeker}, \& {Demory}}]{Hu2024}
{Hu}, R., {Bello-Arufe}, A., {Zhang}, M., {et~al.} 2024, \nat, 630, 609, \dodoi{10.1038/s41586-024-07432-x}

\bibitem[{{Kauahikaua} {et~al.}(2002){Kauahikaua}, {Cashman}, {Clague}, {Champion}, \& {Hagstrum}}]{Kauahikaua2002}
{Kauahikaua}, J., {Cashman}, K., {Clague}, D., {Champion}, D., \& {Hagstrum}, J. 2002, Bull Volcanol, 64, 229, \dodoi{10.1007/s00445-001-0196-8}

\bibitem[{{Kreidberg} {et~al.}(2019){Kreidberg}, {Koll}, {Morley}, {Hu}, {Schaefer}, {Deming}, {Stevenson}, {Dittmann}, {Vanderburg}, {Berardo}, {Guo}, {Stassun}, {Crossfield}, {Charbonneau}, {Latham}, {Loeb}, {Ricker}, {Seager}, \& {Vanderspek}}]{Kreidberg2019}
{Kreidberg}, L., {Koll}, D. D.~B., {Morley}, C., {et~al.} 2019, \nat, 573, 87, \dodoi{10.1038/s41586-019-1497-4}

\bibitem[{{Lichtenberg} \& {Miguel}(2024)}]{Lichtenberg2024}
{Lichtenberg}, T., \& {Miguel}, Y. 2024, arXiv e-prints, arXiv:2405.04057, \dodoi{10.48550/arXiv.2405.04057}

\bibitem[{{Lim} {et~al.}(2023){Lim}, {Benneke}, {Doyon}, {MacDonald}, {Piaulet}, {Artigau}, {Coulombe}, {Radica}, {L'Heureux}, {Albert}, {Rackham}, {de Wit}, {Salhi}, {Roy}, {Flagg}, {Fournier-Tondreau}, {Taylor}, {Cook}, {Lafreni{\`e}re}, {Cowan}, {Kaltenegger}, {Rowe}, {Espinoza}, {Dang}, \& {Darveau-Bernier}}]{Lim2023}
{Lim}, O., {Benneke}, B., {Doyon}, R., {et~al.} 2023, \apjl, 955, L22, \dodoi{10.3847/2041-8213/acf7c4}

\bibitem[{{Logan} {et~al.}(1973){Logan}, {Hunt}, {Salisbury}, \& {Balsamo}}]{Logan1973}
{Logan}, L.~M., {Hunt}, G.~R., {Salisbury}, J.~W., \& {Balsamo}, S.~R. 1973, \jgr, 78, 4983, \dodoi{10.1029/JB078i023p04983}

\bibitem[{{Luque} \& {Pall{\'e}}(2022)}]{Luque&Palle2022}
{Luque}, R., \& {Pall{\'e}}, E. 2022, Science, 377, 1211, \dodoi{10.1126/science.abl7164}

\bibitem[{{Lustig-Yaeger} {et~al.}(2023){Lustig-Yaeger}, {Fu}, {May}, {Ceballos}, {Moran}, {Peacock}, {Stevenson}, {Kirk}, {L{\'o}pez-Morales}, {MacDonald}, {Mayorga}, {Sing}, {Sotzen}, {Valenti}, {Redai}, {Alam}, {Batalha}, {Bennett}, {Gonzalez-Quiles}, {Kruse}, {Lothringer}, {Rustamkulov}, \& {Wakeford}}]{Lustig-Yaeger2023}
{Lustig-Yaeger}, J., {Fu}, G., {May}, E.~M., {et~al.} 2023, Nature Astronomy, 7, 1317, \dodoi{10.1038/s41550-023-02064-z}

\bibitem[{{Mansfield} {et~al.}(2019){Mansfield}, {Kite}, {Hu}, {Koll}, {Malik}, {Bean}, \& {Kempton}}]{Mansfield2019}
{Mansfield}, M., {Kite}, E.~S., {Hu}, R., {et~al.} 2019, \apj, 886, 141, \dodoi{10.3847/1538-4357/ab4c90}

\bibitem[{{Maturilli} \& {Helbert}(2014)}]{Maturilli2014}
{Maturilli}, A., \& {Helbert}, J. 2014, Journal of Applied Remote Sensing, 8, 084985, \dodoi{10.1117/1.JRS.8.084985}

\bibitem[{{Maturilli} {et~al.}(2019){Maturilli}, {Helbert}, \& {Arnold}}]{Maturilli2019}
{Maturilli}, A., {Helbert}, J., \& {Arnold}, G. 2019, in Society of Photo-Optical Instrumentation Engineers (SPIE) Conference Series, Vol. 11128, Infrared Remote Sensing and Instrumentation XXVII, ed. M.~{Strojnik} \& G.~E. {Arnold}, 111280T, \dodoi{10.1117/12.2529266}

\bibitem[{{Maturilli} {et~al.}(2006){Maturilli}, {Helbert}, {Witzke}, \& {Moroz}}]{Maturilli2006}
{Maturilli}, A., {Helbert}, J., {Witzke}, A., \& {Moroz}, L. 2006, \planss, 54, 1057, \dodoi{10.1016/j.pss.2005.12.021}

\bibitem[{{May} {et~al.}(2023){May}, {MacDonald}, {Bennett}, {Moran}, {Wakeford}, {Peacock}, {Lustig-Yaeger}, {Highland}, {Stevenson}, {Sing}, {Mayorga}, {Batalha}, {Kirk}, {L{\'o}pez-Morales}, {Valenti}, {Alam}, {Alderson}, {Fu}, {Gonzalez-Quiles}, {Lothringer}, {Rustamkulov}, \& {Sotzen}}]{May2023}
{May}, E.~M., {MacDonald}, R.~J., {Bennett}, K.~A., {et~al.} 2023, \apjl, 959, L9, \dodoi{10.3847/2041-8213/ad054f}

\bibitem[{{Milliken} {et~al.}(2021){Milliken}, {Hiroi}, {Scholes}, {Slavney}, \& {Arvidson}}]{Milliken2021}
{Milliken}, R.~E., {Hiroi}, T., {Scholes}, D., {Slavney}, S., \& {Arvidson}, R. 2021, in LPI Contributions, Vol. 2654, Astromaterials Data Management in the Era of Sample-Return Missions Community Workshop, 2021

\bibitem[{{Moran} {et~al.}(2023){Moran}, {Stevenson}, {Sing}, {MacDonald}, {Kirk}, {Lustig-Yaeger}, {Peacock}, {Mayorga}, {Bennett}, {L{\'o}pez-Morales}, {May}, {Rustamkulov}, {Valenti}, {Adams Redai}, {Alam}, {Batalha}, {Fu}, {Gonzalez-Quiles}, {Highland}, {Kruse}, {Lothringer}, {Ortiz Ceballos}, {Sotzen}, \& {Wakeford}}]{Moran2023}
{Moran}, S.~E., {Stevenson}, K.~B., {Sing}, D.~K., {et~al.} 2023, \apjl, 948, L11, \dodoi{10.3847/2041-8213/accb9c}

\bibitem[{{Mustard} \& {Hays}(1997)}]{Mustard1997}
{Mustard}, J.~F., \& {Hays}, J.~E. 1997, \icarus, 125, 145, \dodoi{10.1006/icar.1996.5583}

\bibitem[{{Neil} \& {Rogers}(2020)}]{Neil&Rogers2020}
{Neil}, A.~R., \& {Rogers}, L.~A. 2020, \apj, 891, 12, \dodoi{10.3847/1538-4357/ab6a92}

\bibitem[{{Paige} {et~al.}(2010){Paige}, {Foote}, {Greenhagen}, {Schofield}, {Calcutt}, {Vasavada}, {Preston}, {Taylor}, {Allen}, {Snook}, {Jakosky}, {Murray}, {Soderblom}, {Jau}, {Loring}, {Bulharowski}, {Bowles}, {Thomas}, {Sullivan}, {Avis}, {de Jong}, {Hartford}, \& {McCleese}}]{Paige2010}
{Paige}, D.~A., {Foote}, M.~C., {Greenhagen}, B.~T., {et~al.} 2010, \ssr, 150, 125, \dodoi{10.1007/s11214-009-9529-2}

\bibitem[{{Rogers}(2015)}]{Rogers2015}
{Rogers}, L.~A. 2015, \apj, 801, 41, \dodoi{10.1088/0004-637X/801/1/41}

\bibitem[{Rossman \& Ehlmann(2019)}]{Rossman2019}
Rossman, G.~R., \& Ehlmann, B.~L. 2019, Electronic Spectra of Minerals in the Visible and Near-Infrared Regions, Cambridge Planetary Science (Cambridge University Press), 3–20

\bibitem[{{Ruff} {et~al.}(1997){Ruff}, {Christensen}, {Barbera}, \& {Anderson}}]{Ruff1997}
{Ruff}, S.~W., {Christensen}, P.~R., {Barbera}, P.~W., \& {Anderson}, D.~L. 1997, \jgr, 102, 14899, \dodoi{10.1029/97JB00593}

\bibitem[{Salisbury(1993)}]{Salisbury1993}
Salisbury, J.~W. 1993, Mid-Infrared Spectroscopy: Laboratory Data, Cambridge Planetary Science (Cambridge University Press)

\bibitem[{{Salisbury} {et~al.}(1991){Salisbury}, {D'Aria}, \& {Jarosewich}}]{Salisbury1991}
{Salisbury}, J.~W., {D'Aria}, D.~M., \& {Jarosewich}, E. 1991, \icarus, 92, 280, \dodoi{10.1016/0019-1035(91)90052-U}

\bibitem[{{Salisbury} {et~al.}(1994){Salisbury}, {Wald}, \& {D'Aria}}]{Salisbury1994}
{Salisbury}, J.~W., {Wald}, A., \& {D'Aria}, D.~M. 1994, \jgr, 99, 11897, \dodoi{10.1029/93JB03600}

\bibitem[{{Salisbury} \& {Walter}(1989)}]{Salisbury1989}
{Salisbury}, J.~W., \& {Walter}, L.~S. 1989, \jgr, 94, 9192, \dodoi{10.1029/JB094iB07p09192}

\bibitem[{{Seager} \& {Deming}(2009)}]{Seager2009}
{Seager}, S., \& {Deming}, D. 2009, \apj, 703, 1884, \dodoi{10.1088/0004-637X/703/2/1884}

\bibitem[{{Speagle}(2020)}]{Speagle2020}
{Speagle}, J.~S. 2020, \mnras, 493, 3132, \dodoi{10.1093/mnras/staa278}

\bibitem[{Thayer(1934)}]{Thayer1934}
Thayer, T.~P. 1934, PhD thesis, Division of Geological and Planetary Sciences, California Institute of Technology

\bibitem[{{TRAPPIST-1 JWST Community Initiative} {et~al.}(2023){TRAPPIST-1 JWST Community Initiative}, {de Wit}, {Doyon}, {Rackham}, {Lim}, {Ducrot}, {Kreidberg}, {Benneke}, {Ribas}, {Berardo}, {Niraula}, {Iyer}, {Shapiro}, {Kostogryz}, {Witzke}, {Gillon}, {Agol}, {Meadows}, {Burgasser}, {Owen}, {Fortney}, {Selsis}, {Bello-Arufe}, {Bolmont}, {Cowan}, {Dong}, {Drake}, {Garcia}, {Greene}, {Haworth}, {Hu}, {Kane}, {Kervella}, {Koll}, {Krissansen-Totton}, {Lagage}, {Lichtenberg}, {Lustig-Yaeger}, {Lingam}, {Turbet}, {Seager}, {Barkaoui}, {Bell}, {Burdanov}, {Cadieux}, {Charnay}, {Cloutier}, {Cook}, {Correia}, {Dang}, {Daylan}, {Delrez}, {Edwards}, {Fauchez}, {Flagg}, {Fraschetti}, {Haqq-Misra}, {Huang}, {Iro}, {Jayawardhana}, {Jehin}, {Jin}, {Kite}, {Kitzmann}, {Kral}, {Lafreni{\`e}re}, {Libert}, {Liu}, {Mohanty}, {Morris}, {Murray}, {Piaulet}, {Pozuelos}, {Radica}, {Ranjan}, {Rathcke}, {Roy}, {Schwieterman}, {Turner}, {Triaud}, \& {Way}}]{TRAPPIST-1JWSTCommunityInitiative2023}
{TRAPPIST-1 JWST Community Initiative}, {de Wit}, J., {Doyon}, R., {et~al.} 2023, arXiv e-prints, arXiv:2310.15895, \dodoi{10.48550/arXiv.2310.15895}

\bibitem[{{Vanderspek} {et~al.}(2019){Vanderspek}, {Huang}, {Vanderburg}, {Ricker}, {Latham}, {Seager}, {Winn}, {Jenkins}, {Burt}, {Dittmann}, {Newton}, {Quinn}, {Shporer}, {Charbonneau}, {Irwin}, {Ment}, {Winters}, {Collins}, {Evans}, {Gan}, {Hart}, {Jensen}, {Kielkopf}, {Mao}, {Waalkes}, {Bouchy}, {Marmier}, {Nielsen}, {Ottoni}, {Pepe}, {S{\'e}gransan}, {Udry}, {Henry}, {Paredes}, {James}, {Hinojosa}, {Silverstein}, {Palle}, {Berta-Thompson}, {Crossfield}, {Davies}, {Dragomir}, {Fausnaugh}, {Glidden}, {Pepper}, {Morgan}, {Rose}, {Twicken}, {Villase{\~n}or}, {Yu}, {Bakos}, {Bean}, {Buchhave}, {Christensen-Dalsgaard}, {Christiansen}, {Ciardi}, {Clampin}, {De Lee}, {Deming}, {Doty}, {Jernigan}, {Kaltenegger}, {Lissauer}, {McCullough}, {Narita}, {Paegert}, {Pal}, {Rinehart}, {Sasselov}, {Sato}, {Sozzetti}, {Stassun}, \& {Torres}}]{Vanderspek2019}
{Vanderspek}, R., {Huang}, C.~X., {Vanderburg}, A., {et~al.} 2019, \apjl, 871, L24, \dodoi{10.3847/2041-8213/aafb7a}

\bibitem[{{Wachiraphan} {et~al.}(2024){Wachiraphan}, {Berta-Thompson}, {Diamond-Lowe}, {Winters}, {Murray}, {Zhang}, {Xue}, {Morley}, {Rosario-Franco}, \& {Duvvuri}}]{Wachiraphan2024}
{Wachiraphan}, P., {Berta-Thompson}, Z.~K., {Diamond-Lowe}, H., {et~al.} 2024, arXiv e-prints, arXiv:2410.10987, \dodoi{10.48550/arXiv.2410.10987}

\bibitem[{{Weiner Mansfield} {et~al.}(2024){Weiner Mansfield}, {Xue}, {Zhang}, {Mahajan}, {Ih}, {Koll}, {Bean}, {Coy}, {Eastman}, {Kempton}, \& {Kite}}]{WeinerMansfield2024}
{Weiner Mansfield}, M., {Xue}, Q., {Zhang}, M., {et~al.} 2024, \apjl, 975, L22, \dodoi{10.3847/2041-8213/ad8161}

\bibitem[{{Weiss} \& {Marcy}(2014)}]{Weiss2014}
{Weiss}, L.~M., \& {Marcy}, G.~W. 2014, \apjl, 783, L6, \dodoi{10.1088/2041-8205/783/1/L6}

\bibitem[{{Whittaker} {et~al.}(2022){Whittaker}, {Malik}, {Ih}, {Kempton}, {Mansfield}, {Bean}, {Kite}, {Koll}, {Cronin}, \& {Hu}}]{Whittaker2022}
{Whittaker}, E.~A., {Malik}, M., {Ih}, J., {et~al.} 2022, \aj, 164, 258, \dodoi{10.3847/1538-3881/ac9ab3}

\bibitem[{{Wolfgang} \& {Lopez}(2015)}]{Wolfgang2015}
{Wolfgang}, A., \& {Lopez}, E. 2015, \apj, 806, 183, \dodoi{10.1088/0004-637X/806/2/183}

\bibitem[{{Wordsworth} \& {Kreidberg}(2022)}]{Wordsworth2022}
{Wordsworth}, R., \& {Kreidberg}, L. 2022, \araa, 60, 159, \dodoi{10.1146/annurev-astro-052920-125632}

\bibitem[{{Xue} {et~al.}(2024){Xue}, {Bean}, {Zhang}, {Mahajan}, {Ih}, {Eastman}, {Lunine}, {Mansfield}, {Coy}, {Kempton}, {Koll}, \& {Kite}}]{Xue2024}
{Xue}, Q., {Bean}, J.~L., {Zhang}, M., {et~al.} 2024, \apjl, 973, L8, \dodoi{10.3847/2041-8213/ad72e9}

\bibitem[{{Zahnle} \& {Catling}(2017)}]{Zahnle2017}
{Zahnle}, K.~J., \& {Catling}, D.~C. 2017, \apj, 843, 122, \dodoi{10.3847/1538-4357/aa7846}

\bibitem[{{Zhang} {et~al.}(2019){Zhang}, {Chachan}, {Kempton}, \& {Knutson}}]{Zhang2019}
{Zhang}, M., {Chachan}, Y., {Kempton}, E. M.~R., \& {Knutson}, H.~A. 2019, \pasp, 131, 034501, \dodoi{10.1088/1538-3873/aaf5ad}

\bibitem[{{Zhang} {et~al.}(2020){Zhang}, {Chachan}, {Kempton}, {Knutson}, \& {Chang}}]{Zhang2020}
{Zhang}, M., {Chachan}, Y., {Kempton}, E. M.~R., {Knutson}, H.~A., \& {Chang}, W.~H. 2020, \apj, 899, 27, \dodoi{10.3847/1538-4357/aba1e6}

\bibitem[{{Zhang} {et~al.}(2024){Zhang}, {Hu}, {Inglis}, {Dai}, {Bean}, {Knutson}, {Lam}, {Goffo}, \& {Gandolfi}}]{Zhang2024}
{Zhang}, M., {Hu}, R., {Inglis}, J., {et~al.} 2024, \apjl, 961, L44, \dodoi{10.3847/2041-8213/ad1a07}

\bibitem[{{Zhuang} {et~al.}(2023){Zhuang}, {Zhang}, {Ma}, {Jiang}, {Yang}, {Milliken}, \& {Hsu}}]{Zhuang2023}
{Zhuang}, Y., {Zhang}, H., {Ma}, P., {et~al.} 2023, \icarus, 391, 115346, \dodoi{10.1016/j.icarus.2022.115346}

\bibitem[{{Zieba} {et~al.}(2022){Zieba}, {Zilinskas}, {Kreidberg}, {Nguyen}, {Miguel}, {Cowan}, {Pierrehumbert}, {Carone}, {Dang}, {Hammond}, {Louden}, {Lupu}, {Malavolta}, \& {Stevenson}}]{Zieba2022}
{Zieba}, S., {Zilinskas}, M., {Kreidberg}, L., {et~al.} 2022, \aap, 664, A79, \dodoi{10.1051/0004-6361/202142912}

\bibitem[{{Zieba} {et~al.}(2023){Zieba}, {Kreidberg}, {Ducrot}, {Gillon}, {Morley}, {Schaefer}, {Tamburo}, {Koll}, {Lyu}, {Acu{\~n}a}, {Agol}, {Iyer}, {Hu}, {Lincowski}, {Meadows}, {Selsis}, {Bolmont}, {Mandell}, \& {Suissa}}]{Zieba2023}
{Zieba}, S., {Kreidberg}, L., {Ducrot}, E., {et~al.} 2023, \nat, 620, 746, \dodoi{10.1038/s41586-023-06232-z}

\end{thebibliography}
\bibliographystyle{aasjournal}

\appendix
\section{Sample Information}

\begin{table}[p]
\centering
\rotatebox{90}{
\begin{minipage}{\textheight}
\caption{Origins of Samples}
\small
\label{tab:sample_info}
\begin{tabular}{llll}
\hline
Sample & Origin & Source & Notes \cr
\hline
dunite xenolith & Hualalai Volcano, HI & Paul Asimow & \citet{Kauahikaua2002} \cr
basalt w/ phenocrysts & Hualalai Volcano, HI & Paul Asimow & -- \cr
olivine clinopyroxenite EG-19-63\textsuperscript{a}  & Emigrant Gap, CA & Juliet Ryan-Davis & Ryan-Davis et al. (in prep.) \cr
olivine clinopyroxenite EG-19-70\textsuperscript{a} & Emigrant Gap, CA & Juliet Ryan-Davis & Ryan-Davis et al. (in prep.) \cr
basaltic andesite & Eastern Lau Spreading Center & Paul Asimow & Cruise DOI: 10.7284/901688 \cr
K1919\textsuperscript{a} & Kilauea, HI & Caltech collection & \citet{Anderson2017} \cr
olivine gabbronorite EG-19-68\textsuperscript{a} & Emigrant Gap, CA  & Juliet Ryan-Davis &  Ryan-Davis et al. (in prep.) \cr
andesite STM-101\textsuperscript{a} & Santiam, OR & Caltech collection & \citet{Thayer1934} \cr
dalmatian granite & Unknown & T\&L Granite Countertop Warehouse & Located in El Monte, CA \cr
orlando gold granite & Unknown & T\&L Granite Countertop Warehouse & Located in El Monte, CA \cr
\hline
\end{tabular}
\begin{flushleft}
\small{
\textsuperscript{a}Naming scheme adopted from source.
}
\end{flushleft}
\end{minipage}
}
\end{table}


\begin{table}[p]
\centering
\rotatebox{90}{
\begin{minipage}{\textheight}
\caption{Chemical compositions of samples (wt\%). The detection limit for all of the samples is 0.01\% except for \ce{MnO} (0.005\%) and \ce{TiO2} (0.001\%).}
\small
\begin{tabular}{l|rrrrrrrrrrrr}
\hline
Sample Name & \ce{SiO2} & \ce{Al2O3} & \ce{Fe2O3}(T) & \ce{MnO} & \ce{MgO} & \ce{CaO} & \ce{Na2O} & \ce{K2O} & \ce{TiO2} & \ce{P2O5} & LOI\textsuperscript{a} & Total \\
\hline
hematite & 2.15 & 0.54 & 96.45 & 0.04 & 0.05 & 0.19 & 0.02 & $<$0.01 & 0.08 & 0.13 & 0.46 & 100.10 \\
dunite xenolith & 39.04 & 0.19 & 12.86 & 0.16 & 49.28 & 0.24 & 0.04 & $<$0.01 & 0.03 & $<$0.01 & -1.01 & 100.80 \\
basalt w/ phenocrysts & 44.50 & 10.12 & 13.13 & 0.17 & 15.78 & 11.07 & 1.57 & 0.29 & 1.86 & 0.21 & 0.83 & 99.54 \\
olivine clinopyroxenite EG-19-63 & 47.10 & 1.72 & 9.07 & 0.18 & 24.06 & 16.04 & 0.19 & 0.01 & 0.16 & $<$0.01 & 0.44 & 98.96 \\
olivine clinopyroxenite EG-19-70 & 47.58 & 1.71 & 11.26 & 0.20 & 19.62 & 16.20 & 0.18 & 0.01 & 0.25 & $<$0.01 & 1.84 & 98.87 \\
basaltic andesite & 49.94 & 12.77 & 13.94 & 0.18 & 8.53 & 10.48 & 2.22 & 0.39 & 2.32 & 0.23 & -0.88 & 100.10 \\
K1919 basalt & 50.65 & 13.52 & 12.79 & 0.17 & 6.83 & 11.57 & 2.28 & 0.52 & 2.80 & 0.28 & -1.39 & 100.50 \\
olivine gabbronorite EG-19-68 & 51.05 & 13.32 & 10.32 & 0.18 & 11.65 & 11.33 & 1.96 & 0.35 & 0.39 & $<$0.01 & -0.13 & 100.40 \\
andesite STM-101 & 52.82 & 16.02 & 12.10 & 0.20 & 4.42 & 8.43 & 3.30 & 0.80 & 1.75 & 0.31 & 0.73 & 100.90 \\
dalmatian granite & 67.51 & 16.35 & 3.56 & 0.05 & 0.73 & 2.41 & 2.49 & 6.32 & 0.40 & 0.16 & 0.63 & 100.60 \\
orlando gold granite & 73.30 & 15.36 & 1.24 & 0.03 & 0.10 & 1.08 & 3.88 & 5.31 & 0.01 & $<$0.01 & 0.26 & 100.60 \\
\hline
\end{tabular}
\begin{flushleft}
\small{
\textsuperscript{a}Loss on Ignition: represents the amount of organic matter that was in the sample
}
\end{flushleft}
\label{table:chemistry}
\end{minipage}
}
\end{table}


\begin{table}[p]
\centering
\rotatebox{90}{
\begin{minipage}{\textheight}
\caption{Mineral abundances of samples in wt\%.}
\small
\begin{tabular}{l|rrrrrrrrrrrrrrrr}
\hline
Sample Name & \rotatebox{90}{Plagioclase} & \rotatebox{90}{K feldspar} & \rotatebox{90}{Quartz} & \rotatebox{90}{Augite/Diopside} & \rotatebox{90}{Forsterite, ferroan} & \rotatebox{90}{Amphibole} & \rotatebox{90}{Enstatite} & \rotatebox{90}{Serpentine} & \rotatebox{90}{Talc} & \rotatebox{90}{Muscovite/Illite} & \rotatebox{90}{Biotite} & \rotatebox{90}{Chlorite} & \rotatebox{90}{Ilmenite} & \rotatebox{90}{Magnetite} & \rotatebox{90}{Hematite} & \rotatebox{90}{Amorphous} \\
\hline
hematite & n.d. & n.d. & 3.4 & n.d. & n.d. & n.d. & n.d. & n.d. & n.d. & n.d. & n.d. & n.d. & n.d. & n.d. & 80.8 & 15.8 \\
dunite xenolith & n.d. & n.d. & n.d. & n.d. & 100.0 & n.d. & n.d. & n.d. & n.d. & n.d. & n.d. & n.d. & n.d. & n.d. & n.d. & n.d. \\
basalt w/ phenocrysts & 47.8 & n.d. & n.d. & 34.8 & n.d. & n.d. & n.d. & n.d. & n.d. & n.d. & n.d. & n.d. & trace & n.d. & n.d. & 17.4 \\
olivine clinopyrox. EG-19-63 & 2.0 & n.d. & n.d. & 66.9 & 26.9 & 2.2 & n.d. & 2.0 & n.d. & n.d. & n.d. & n.d. & n.d. & n.d. & n.d. & n.d. \\
olivine clinopyrox. EG-19-70 & 34.4 & n.d. & n.d. & 34.9 & 23.9 & n.d. & n.d. & n.d. & n.d. & n.d. & n.d. & n.d. & n.d. & n.d. & n.d. & 6.8 \\
basaltic andesite & 21.0 & n.d. & n.d. & 35.2 & 7.4 & n.d. & n.d. & n.d. & n.d. & n.d. & n.d. & n.d. & n.d. & n.d. & n.d. & 36.4 \\
K1919 basalt & 47.8 & n.d. & n.d. & 34.8 & n.d. & n.d. & n.d. & n.d. & n.d. & n.d. & n.d. & n.d. & trace & n.d. & n.d. & 17.4 \\
olivine gabbronor. EG-19-68 & 69.9 & n.d. & 6.3 & 19.7 & n.d. & n.d. & n.d. & n.d. & n.d. & n.d. & n.d. & n.d. & 1.5 & 2.6 & n.d. & n.d. \\
andesite STM-101 & 44.9 & n.d. & n.d. & 24.3 & n.d. & n.d. & 29.8 & n.d. & n.d. & 1.0 & n.d. & n.d. & n.d. & n.d. & n.d. & n.d. \\
dalmatian granite & 25.7 & 34.1 & 29.7 & n.d. & n.d. & n.d. & n.d. & n.d. & n.d. & n.d. & 8.7 & 1.7 & n.d. & n.d. & n.d. & n.d. \\
orlando gold granite & 37.0 & 28.1 & 28.9 & n.d. & n.d. & n.d. & n.d. & n.d. & n.d. & 2.0 & n.d. & n.d. & n.d. & n.d. & n.d. & 4.0 \\
\hline
\label{table:mineralogy}
\end{tabular}
\end{minipage}
}
\end{table}

\section{Determining the Optimal Redistribution Factor} \label{appendix: f}
\begin{figure}[ht!]
    \centering
    \includegraphics[width=0.45\textwidth]{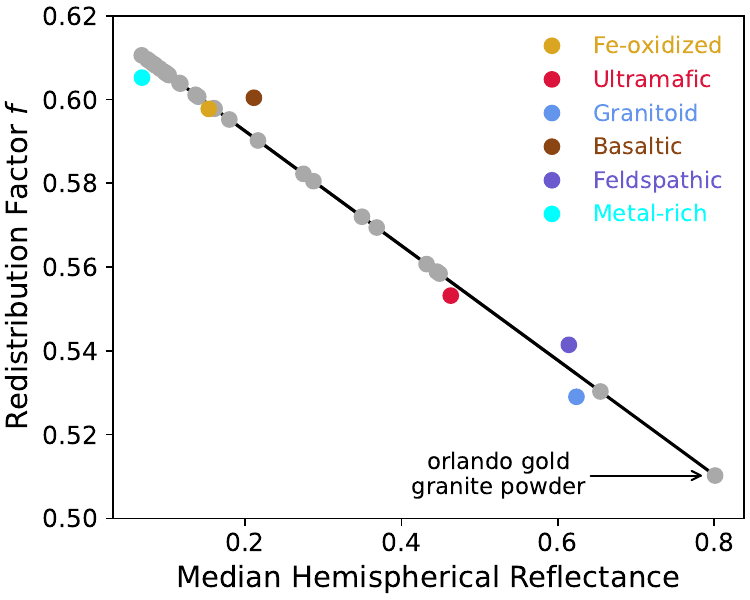}
    \caption{The best-fit $f$ values for the surfaces in \citet{Hu2012} as a function of median hemispherical reflectance taken over a wavelength range corresponding to $85\%$ of the integrated flux of a \SI{3000}{\kelvin} host star. The best-fit linear function used to estimate the $f$ values for the samples in our expanded spectral library is shown as a solid black line. Each new sample is shown as a grey marker. The range in median hemispherical reflectance for the linear function corresponds to the range of reflectances of the samples in the library.}
    \label{fig:f}
\end{figure}

To ensure that our 1D models capture the temperature gradient on the dayside surface of a planet, we used a redistribution factor $f$ in our energy balance equation as described in \S\ref{sec:modeling}. We estimated the best-fit $f$ for each surface in \citet{Hu2012} by scaling $f$ until our 1D models matched the \citet{Hu2012} 2D models from \SIrange{0}{25}{\micro\meter}, which summed the flux contributions from patches with different temperatures across the dayside. The best-fit $f$ values for a planet with an equilibrium temperature of \SI{1000}{\kelvin} as a function of median hemispherical reflectance are shown as colored points in Figure~\ref{fig:f}, and scale approximately linearly with the median hemispherical reflectance of each sample. We repeated this fitting for a narrower wavelength range, \SIrange{5}{12}{\micro\meter}, and found that the differences in the generated models were negligible.

We performed this fitting for a range of equilibrium temperatures and found that the best-fit $f$ is weakly dependent on equilibrium temperature. Specifically, colder planets were better matched by larger $f$ values. We found that the difference in best-fit $f$ is negligible relative to the expected observational precision for \textit{JWST}, and therefore opted to use the $f$ value corresponding to a planet with an equilibrium temperature of \SI{1000}{\kelvin} for all models. We quantified the impact of this assumption by calculating the emission spectrum for a planet with a cooler equilibrium temperature of \SI{400}{\kelvin}, and found that the difference between the temperature-dependent $f$ model and fixed $f$ model at \SI{9}{\micro\meter} (in the middle of the Si-O stretching feature) for every surface ranged from 0\% to 5.3\%. Taking the Trappist-1 system as our test case, we find that even when combining multiple eclipse observations in a broad photometric bandpass the \textit{JWST} measurement precision for planets cooler than 600~K exceeds 10\% \citep{Greene2023,Zieba2023}. We therefore do not expect these small differences to affect our ability to proficiently model the data, and leave them out of the default model implementation in \texttt{PLATON}. However, we nonetheless include the temperature-dependent values in Table~\ref{tab:f_equilibrium_temps} in case they are useful for future studies. 

\begin{figure}[ht!]
    \centering
    \includegraphics[width=0.85\textwidth]{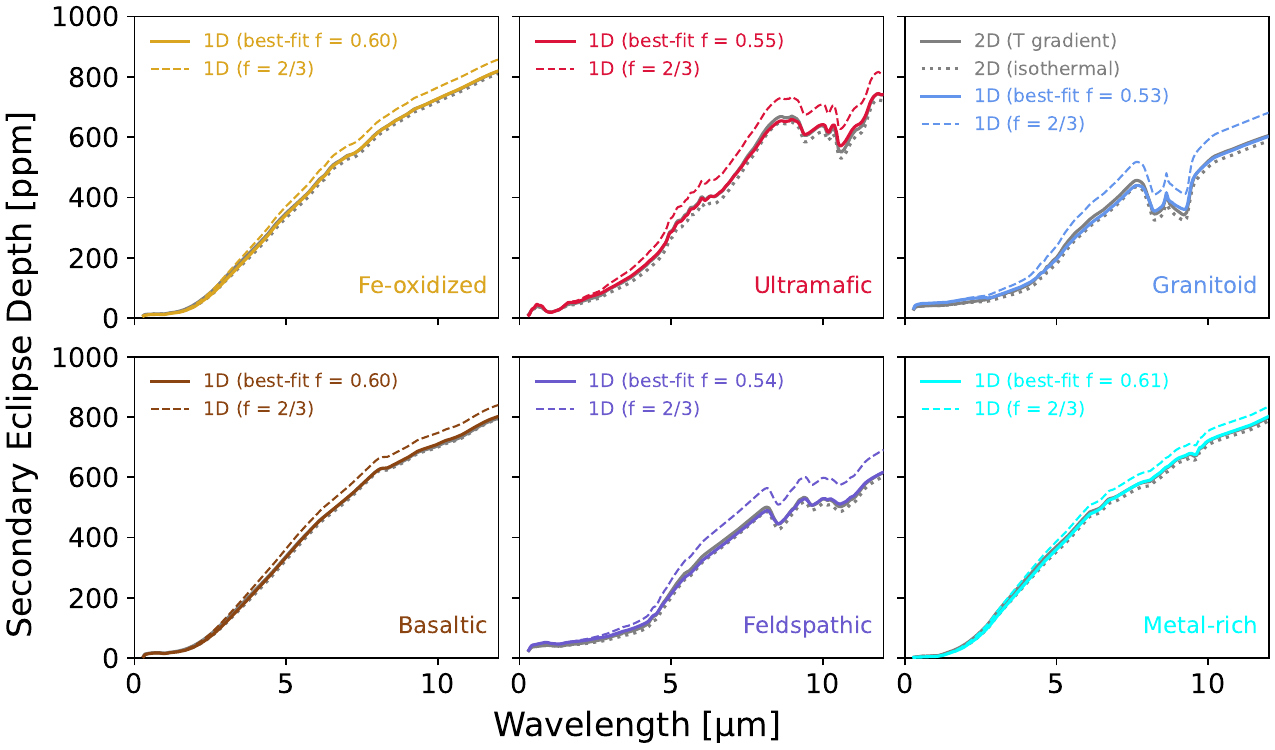}
    \caption{The 1D models described in this work compared to the 2D models of the surfaces in \citet{Hu2012} for LHS 3844 b system parameters. We show our 1D models for the best-fit redistribution factor and f=2/3, the typical value used in the literature. For full comparison, we plot the full 2D model and the isothermal 2D model, i.e., a uniform temperature across the dayside of the planet, from \citet{Hu2012} in each panel.}
    \label{fig:compare}
\end{figure}

We show how our 1D models compare to the 2D models from \citet{Hu2012} in Figure~\ref{fig:compare} for LHS 3844 b system parameters. Our 1D model spectra with a uniform dayside temperature are very similar to the 2D model spectra with, and without, a temperature gradient. They are not identical because the 2D models additionally incorporate angle of emission. We also show what the spectra look like with f=2/3, and see that this assumption generally leads to an overestimation of the planet flux. This overestimation is the greatest for the most reflective surfaces.

\begin{table}
\centering
\small
\begin{tabular}{c|ccccccc}
\hline
Teq [K] & Fe-oxidized & Ultramafic & Granitoid & Basaltic & Feldspathic & Metal-rich \\
\hline
1100 & 0.5971 & 0.5269 & 0.5120 & 0.5965 & 0.5220 & 0.6066 \\
1050 & 0.5954 & 0.5336 & 0.5156 & 0.5968 & 0.5285 & 0.6068 \\
1000 & 0.5947 & 0.5354 & 0.5172 & 0.5971 & 0.5300 & 0.6070 \\
950 & 0.5939 & 0.5371 & 0.5189 & 0.5974 & 0.5315 & 0.6072 \\
900 & 0.5931 & 0.5388 & 0.5206 & 0.5977 & 0.5330 & 0.6074 \\
850 & 0.5923 & 0.5406 & 0.5223 & 0.5980 & 0.5345 & 0.6076 \\
800 & 0.5915 & 0.5423 & 0.5239 & 0.5983 & 0.5360 & 0.6078 \\
750 & 0.5936 & 0.5440 & 0.5256 & 0.5986 & 0.5375 & 0.6080 \\
700 & 0.5936 & 0.5458 & 0.5273 & 0.5988 & 0.5390 & 0.6082 \\
650 & 0.5937 & 0.5475 & 0.5290 & 0.5991 & 0.5405 & 0.6084 \\
600 & 0.5938 & 0.5492 & 0.5307 & 0.5994 & 0.5420 & 0.6086 \\
550 & 0.5984 & 0.5634 & 0.5305 & 0.5991 & 0.5457 & 0.6060 \\
500 & 0.5951 & 0.5626 & 0.5314 & 0.6022 & 0.5457 & 0.6027 \\
450 & 0.5966 & 0.5656 & 0.5292 & 0.6029 & 0.5446 & 0.6073 \\
400 & 0.5991 & 0.5658 & 0.5310 & 0.6034 & 0.5467 & 0.6083 \\
350 & 0.5998 & 0.5675 & 0.5311 & 0.6046 & 0.5469 & 0.6098 \\
300 & 0.6031 & 0.5701 & 0.5325 & 0.6077 & 0.5492 & 0.6125 \\
\hline
\end{tabular}
\caption{The redistribution factor $f$ as a function of equilibrium temperature $T_\mathrm{eq}$ in kelvin for the surfaces in \citet{Hu2012}.}
\label{tab:f_equilibrium_temps}
\end{table}

Rather than calculating the full 2D Hu et al. models for the new samples presented in this work in order to estimate their respective $f$ values, we instead leveraged the approximately linear relationship between the best-fit $f$ values as a function of median hemispherical reflectance. We fit the points in Figure~\ref{fig:f} with a linear function and found coefficients of $m = -0.137$ and $b = 0.620$. We then used this relationship to obtain $f$ values for our new samples. We test the validity of this approach using the most reflective rock and texture combination in the library, which is the orlando gold granite powder (median hemispherical reflectance of 0.80).  This material is more reflective than any of the surfaces in \citet{Hu2012}. Nonetheless, our linear relation predicts that its corresponding $f$ value is 0.51. This result is reassuring as the value of $f$ should always lie between $1/2$ and $2/3$ \citep{Hansen2008}.





\end{document}